\documentclass [11pt,a4paper]{article}
\usepackage{jcappub}

\usepackage[latin1,utf8]{inputenc}
\usepackage{graphicx}
\usepackage{color}
\usepackage{amsfonts,amsthm,amssymb}
\usepackage{bm,bbm}
\usepackage[OT2,OT1]{fontenc}
\usepackage{tikz}
\usepackage{pifont} 
\usepackage{changepage} 
\usepackage{ulem}

\newcommand\cyr{%
\renewcommand\rmdefault{wncyr}%
\renewcommand\sfdefault{wncyss}%
\renewcommand\encodingdefault{OT2}%
\normalfont
\selectfont}
\DeclareTextFontCommand{\textcyr}{\cyr}

\newcommand{\be}{\begin{equation}}
\newcommand{\ee}{\end{equation}}
\newcommand{\ba}{\begin{eqnarray}}
\newcommand{\ea}{\end{eqnarray}}

\renewcommand{\texttt}{{}}

\def\bs{\begin{subequations}}
\def\es{\end{subequations}}
\def\a{\alpha}
\def\b{\beta}
\def\de{\delta}
\def\De{\Delta}
\def\g{\gamma}
\def\G{\Gamma}
\def\la{\lambda}
\def\k{\kappa}
\def\e{\epsilon}
\def\ve{\varepsilon}
\def\Om{\Omega}
\def\om{\omega}
\def\G{\Gamma}
  
\def\s{\sigma}

\def\vp{\varphi}
\def\N{\nabla}

\def\cF{\mathcal{F}}
\def\cG{\mathcal{G}}
\def\cH{\mathcal{H}}

\def\cJ{\mathcal{J}}
\def\cK{\mathcal{K}}

\def\cP{\mathcal{P}}

\def\cS{\mathcal{S}}

\def\cV{\mathcal{V}}

\def\ds{d_{\rm S}}
\def\dh{d_{\rm H}}
\def\dw{d_{\rm W}}

\def\p{\partial}

\def\B{\Box}

\newcommand{\Eq}[1]{(\ref{#1})}

\def\cob{\color{blue}}
\newcommand{\book}[5]{\emph{#1}, #2, #3, #4 (#5)}
\newcommand{\books}[4]{\emph{#1}, #2, #3 (#4)}
\newcommand{\oarX}[1]{\href{http://arxiv.org/abs/#1}{{\ttfamily\cob arXiv:#1}}}
\newcommand{\arX}[1]{\href{http://arxiv.org/abs/#1}{{\ttfamily\cob arXiv:#1}}}
\newcommand{\doin}[6]{\href{http://dx.doi.org/#1}{{\cob {\it #2} {\bf #3 #4} (#6) #5}}}
\newcommand{\doinn}[5]{\href{http://dx.doi.org/#1}{{\cob {\it #2} {\bf #3} (#5) #4}}}
\newcommand{\doij}[5]{\href{http://dx.doi.org/#1}{{\cob {\it #2} {\bf #3} (#5) #4}}}

\newcommand{\ndoinn}[5]{\href{#1}{{\cob {\it #2} {\bf #3} (#5) #4}}}
\newcommand{\procsinm}[5]{in \emph{#1}, #2 eds., #3, #4 (#5)}

\newcommand{\tia}[1]{\textit{#1},}
\newcommand{\boxd}[1]{\boxed{\phantom{\Biggl(}#1\phantom{\Biggl)}}}
\def\Pl{{\rm Pl}}
\def\lp{\ell_\Pl}
\def\tp{t_\Pl}
\def\ep{E_\Pl}
\def\rme{e}
\def\rmd{d}
\def\rmi{i}
\def\dpl{\delta_{\rm inv}}
\def\geq{\geqslant}
\def\leq{\leqslant}
\definecolor{darkgreen}{RGB}{0,127,50}

\begin{document}


\title{Quantum gravity and gravitational-wave astronomy}

\author[a]{Gianluca Calcagni,}
\affiliation[a]{Instituto de Estructura de la Materia, CSIC, Serrano 121, 28006 Madrid, Spain}
\emailAdd{g.calcagni@csic.es}

\author[b,c]{Sachiko Kuroyanagi,}
\affiliation[b]{Department of Physics, Nagoya University, Chikusa, Nagoya 464-8602, Japan}
\affiliation[c]{Instituto de F\'isica Te\'orica UAM-CSIC, Universidad Auton\'oma de Madrid, \\Cantoblanco, 28049 Madrid, Spain}
\emailAdd{skuro@nagoya-u.jp}

\author[d]{Sylvain Marsat,}
\affiliation[d]{APC, AstroParticule et Cosmologie, Universit\'{e} Paris Diderot, CNRS/IN2P3, CEA/Irfu, Observatoire de Paris, Sorbonne Paris Cit\'{e}, 10, rue Alice Domon et L\'{e}onie Duquet 75205 PARIS Cedex 13, France}
\emailAdd{sylvain.marsat@aei.mpg.de}

\author[e]{Mairi Sakellariadou,}
\affiliation[e]{Theoretical Particle Physics and Cosmology Group, Physics Department, King's College London, University of London, Strand, London WC2R 2LS, United Kingdom}
\emailAdd{mairi.sakellariadou@kcl.ac.uk}

\author[f]{Nicola Tamanini,}
\affiliation[f]{Max-Planck-Institut f\"ur Gravitationsphysik, Albert-Einstein-Institut, Am Mühlenberg 1,
14476 Potsdam-Golm, Germany}
\emailAdd{nicola.tamanini@aei.mpg.de}

\author[g]{Gianmassimo Tasinato}
\affiliation[g]{Department of Physics, Swansea University, Swansea, SA2 8PP, UK}
\emailAdd{g.tasinato2208gmail.com}

\abstract{We investigate possible signatures of quantum gravity which could be tested with current and future gravitational-wave (GW) observations. In particular, we analyze how quantum gravity can influence the GW luminosity distance, the time dependence of the effective Planck mass and the instrumental strain noise of interferometers. Using both model-dependent and model-independent formul\ae, we show that these quantities can encode a non-perturbative effect typical of all quantum-gravity theories, namely the way the dimension of spacetime changes with the probed scale. Effects associated with such dimensional flow might be tested with GW observations and constrained significantly in theories with a microscopically discrete spacetime geometry, more strongly than from propagation-speed constraints. Making use of public LIGO data as well as of a simulated higher-redshift LISA source, we impose the first, respectively, actual and mock constraints on quantum-gravity parameters affecting the GW luminosity distance and discuss specific theoretical examples. If also the Newtonian potential is modified but light geodesics are not, then solar-system bounds may be stronger than GW ones. Yet, for some theories GW astronomy can give unique information not available from solar-system tests.}

\maketitle


\section{Introduction}\label{intro}

Since the dawn of the era of gravitational wave (GW) astronomy opened up by the results obtained by the LIGO-Virgo interferometers \cite{Ab16a,Ab16b,Ab16c,Ab17a,Ab17b}, the interest on how to test new physics with the observations of massive gravitational systems have been mounting at an accelerating pace. In particular, several studies exploring how GWs can and will constrain dark-energy models and quantum-gravity theories appear frequently in the literature. In parallel, activity on model-independent parametrizations and the quest for GW physical observables have reached almost febrile levels, with the aim to optimize the information one will gather when enough GW sources will be available in our catalogs. There is still much to understand but some parameters seem especially promising. The luminosity distance $d_L$ and the time variation of the effective Planck mass are two of them.

Standard general relativity (GR) perfectly explains our current observations \cite{Callister:2017ocg,Ab17b,Abbott:2018utx} but one cannot exclude the possibility that GW astronomy could be the window theories beyond GR have been waiting for. For instance, the landscape of quantum gravities (e.g., \cite{Ori09,Fousp,CQC,MiTr2}) is populated by a number of top-down candidates, each with its own characteristic assumptions and predictions (or lack of them). It is generally accepted that some of them could indeed unify quantum mechanics and gravity in a consistent way, while offering a rich phenomenology prone to experimental verification. Having seen that the cosmic microwave background can discriminate among models of the early universe \cite{P18I}, late-time constraints such as those coming from GWs are the next natural direction where to look into. 

Quantum gravity (QG) and modified gravity can affect both the generation (see, e.g., \cite{YYP,Kobakhidze:2016cqh,BYY,TaYa,Mas18,Giddings:2019ujs}) as well as the propagation of gravitational waves. In what follows, we discuss only the latter. More precisely, we will analyze the luminosity distance and various definitions of effective Planck masses in the context of quantum gravity. In this paper, we will concentrate on the dynamics of the spin-2 (transverse traceless) field $h_{ij}$ and discuss experimental constraints on the effective Lagrangian for this mode. The trace of $h_{\mu\nu}$ is a scalar mode but its dynamics is affected by the same kinetic term of the graviton and, therefore, we will also discuss how the Newton potential may be modified in these theories. Apart from this exception, which entails subtleties we will discuss later, we will ignore any modification on matter (spin-0 and spin-$1/2$ particles) and on gauge fields (spin-1) that could arise in quantum gravity. As restrictive as this assumption may sound, it is prompted by the fact that it is generally difficult to give a unique prescription on QG constraints on all sectors. QG modifications of the Standard Model of particle physics are strongly model dependent while, given the specific but general features of the gravitational interaction, in the gravitational sector there are universal phenomena such as dimensional flow that allow us to make reasonably general statements and place portable observational constraints. Therefore, we will only consider QG models, or special regimes of QG theories, where only the gravitational sector is modified, at least at large scales. 

In some cases, tight constraints from the matter sector exist for quantum gravity (e.g., Lorentz-symmetry violations \cite{CoGl2,KoMe} and bounds on time delay of photons from gamma-ray bursts and other distant sources \cite{Fermi,HESS,Vas13}; see also question 59 of \cite{revmu}) but, in general, they are independent of and complementary to those one can obtain from GWs \cite{qGW} because either they span a different portion in the parameter space of the theories considered here, or they regard different models altogether (for instance, extensions of the Standard Model with explicit Lorentz-violating operators \cite{CoGl2,KoMe} not appearing in our QG list, or photon time delays due to dispersion relations \cite{HESS} not typical of our most promising candidates). Thus, as we will remark again in section \ref{compa}, information from GWs may still be of value for those theories where matter and particle-physics constraints exist, while for theories where such bounds are not yet available (as in the majority of the models considered here) it may be regarded as a first step to connect theory and phenomenology.

Given these premises, we can summarize our results on GWs as follows.
\begin{enumerate}
\item Keeping the discussion as general as possible, we will show that quantum gravity affects the propagation and observation of GWs via global non-perturbative effects on the geometry of spacetime. Formul\ae\ expressing the luminosity distance and the time variation of the Planck mass (controlled by a parameter dubbed $\a_M$ or $\nu$) in terms of spacetime dimensions will be given. The treatment will be model-independent and will encompass several quantum gravities, albeit not all.\footnote{String field theory and non-local quantum gravity are not included, for a technical reason that will become clear later.} Theories to which we will apply our fomul\ae\ are Stelle gravity \cite{Ste77,Ste78} (non-unitary, but still a classic example of renormalizability), the low-energy limit of string theory \cite{Pol98,BBSb,Zwi09}, asymptotic safety \cite{Wei79,Reu1,NiR,Nie06,CPR,Lit11,RSnax}, causal dynamical triangulations \cite{AmJ,AJL4,AJL5,lol08,AJL8,AGJL4,CoJu,CoDo}, non-commutative $\k$-Minkowski spacetime \cite{Sza01,ADKLW,Ben08,ArTr}, Ho\v{r}ava--Lifshitz gravity \cite{Hor09,Hor3,HoMe}, Padmanabhan's non-local field theory near the horizon of black holes \cite{Pad98,Pad99,ArCa1}, and the continuum limit of quantum gravities with discrete pre-geometries: group field theory (GFT) \cite{Fre05,Ori09,BaO11,Fousp,Ori13,GiSi}, spin foams \cite{Per03,Rov10,Per13} and loop quantum gravity (LQG) \cite{rov07,thi01}.
\item Interestingly, our formalism is flexible enough to be applied to other theories not motivated by quantum gravity whenever the dispersion relation of tensor modes or the effective Planck mass are modified. This class includes beyond-GR scenarios such as scalar-tensor theories (see, e.g., \cite{Lan17,KaTs} for reviews). A novel geometric interpretation of these parameters can enrich our understanding and classification of these cases. In the case of beyond-GR scenarios such scalar-tensor systems, the varying-Planck-mass parameter $\a_M=\nu$ can be reinterpreted geometrically as the difference between the topological and the effective Hausdorff dimension of spacetime.
\item As far as observations of standard sirens are concerned, we will make use of public data on the neutron-star merging event GW170817 from the LIGO-Virgo collaboration, and of a simulated super-massive black-hole merging process emitting GWs in the LISA frequency band,  to place the first model-independent observational (real and simulated) constraints on dimensional flow in quantum gravity. To simplify the final result in three points:
\begin{itemize}
\item The spectral dimension of spacetime in the infrared (IR; scales where QG corrections are very small or negligible) is equal to the topological dimension $D=4$, for all theories, by construction. In this regime, GR is recovered.
\item As expected on general grounds, the spectral dimension in the ultraviolet (UV; scales where QG corrections are important) is virtually unconstrained. The deep microscopic regime of the QGs considered here does not affect the propagation of GWs. In other words, we confirm that, typically, GW observations fail to give relevant constraints on many UV-modified quantum-gravity models.
\item However, since dimensional flow is a global characteristic of spacetime, a ``tail'' of QG effects survives at mesoscopic scales (intermediate between the UV and the IR). In this regime, observations imply that the spectral dimension cannot deviate from $D=4$ by more than about 2\%, if the Hausdorff dimension of position and momentum space are close enough to four:
\be\label{1}
\frac{\ds^{\rm meso}-4}{4}<0.02\,.
\ee
\end{itemize}
In particular:
\begin{itemize}
\item Some models are unobservable in any regime (IR, mesoscopic or UV) and luminos\-i\-ty-distance corrections at the Planck scale are negligible in astrophysical processes. Examples are all quantum gravities or beyond-GR models where the only non-perturbative effect is a change of the Hausdorff dimension of spacetime with the probed scale. Still, the strain noise of GW interferometers can constrain the UV Hausdorff dimension of spacetime effectively.
\item Several other quantum gravities also have a varying spectral dimension, but still the effect on the luminosity distance is negligible in all regimes: causal dynamical triangulations, $\k$-Minkowski spacetime, Stelle gravity, the low-energy limit of string theory, asymptotic safety, Ho\v{r}ava--Lifshitz gravity and Padmanabhan's non-local model.
\item Depending on the quantum state giving rise to the background geometry, the group of models GFT/spin foams/LQG may survive in the detectability window of LIGO-Virgo, LISA and third-generation interferometers and they may give rise to non-negligible deviations from the standard luminosity distance in the mesoscopic regime. This happens because the spectral dimension reaches the IR limit from above, which can produce a conspicuous near-IR tail of QG effects.
\item Even in the latter candidates, some semi-classical states are such that corrections are not Planck suppressed, but they are ruled out non-trivially by combining luminosity-distance with propagation-speed constraints. An example is the generic semi-classical state of the effective-dynamics approach of loop quantum cosmology, viable during inflation but completely reduced to a purely GR description (negligible quantum corrections) at late times.
\end{itemize}
\end{enumerate}
Table \ref{tab0} collects these results in a bird's eye view.
\begin{table}
\centering
\begin{tabular}{l|cc|cc}\hline
																							& \multicolumn{2}{c|}{From $d_L$ data} & \multicolumn{2}{c}{From strain noise} \\\hline
																							&	UV & meso				&	UV & meso		\\\hline\hline
GFT/spin foams/LQG 										 				& 	 & \ding{51}							  & \ding{51} &											\\
{\small\qquad (Effective-dynamics LQC)}							& 	 &  											  &        		& 								 	 	\\
Causal dynamical triangulations (phase C)     & 	 &  											  &        		& 								 	 	\\
$\k$-Minkowski (c.i.m.) 											&  	 & 				      					& \ding{51}	& 										\\
$\k$-Minkowski (o.m.)   											&			&  											  &        		& 								 	 	\\
Stelle gravity  														  & 					&  											  &        		& 								 	 	\\
String theory (low-energy limit)							& 					&  											  &        		& 								 	 	\\
Asymptotic safety															& 					&  											  &        		& 								 	 	\\
Ho\v{r}ava--Lifshitz gravity									& 					&  											  &        		& 								 	 	\\
Padmanabhan's non-local model									& 					&  											  &        		& 								 	 	\\\hline\hline
\end{tabular}
\caption{\label{tab0} Observability in GW data of the ultraviolet regime (UV) or the mesoscopic regime (meso) of various theories of quantum gravity with $D=4$ and for string theory in any target-spacetime dimension. A tick indicates that the theory \emph{might} give a detectable signal, while empty cells correspond to theories which \emph{cannot} produce such a signal. In the case of non-commutative $\k$-Minkowski spacetime, ``o.m.'' and ``c.i.m.'' stand for, respectively, ordinary and cyclic-invariant spacetime measure.}
\end{table}

The plan of the paper is as follows. In section \ref{gr1}, we review the definition of luminosity distance, while in section \ref{gwa} we derive the dependence of the gravitational-wave amplitude $h$ on $d_L$ in GR in $D$ dimensions, both reviewing a known calculation \cite{CDL} and proposing a shortcut based on a dimensional argument. The dimensional argument leading to the $D$-dimensional extension of standard luminosity-distance relations will be very instructive about the way we will approach the problem in quantum gravity. In section \ref{para}, parameters and observables relating cosmological theories beyond GR, standard candles and standard sirens are reviewed and new parameters are introduced. Section \ref{qgs} reviews the concepts of Hausdorff and spectral dimension in quantum gravity, and contains  few examples. We will also calculate a scaling parameter (dubbed $\G$) that will appear in the GW amplitude. Section \ref{ludis} is devoted to the study of the GW amplitude $h$ and its relation with the luminosity distance in quantum gravity. The imprint of theories where the Hausdorff dimension of spacetime is scale-dependent is considered in section \ref{ludis1}, while theories where also the spectral dimension varies are discussed in section \ref{ludis2} in a unified, model-independent way. An application of these results to scalar-tensor theories is commented upon in section \ref{dhost}, where the form of the action is given a geometric interpretation. Using data from the binary neutron-star event GW170817 and simulations of a super-massive black-hole merging, in section \ref{nums} we place constraints on quantum-gravity modifications of the luminosity distance, encoded in two geometric parameters (a characteristic length scale $\ell_*$ and a parameter $\g$ combining various definitions of spacetime dimension). In section \ref{othe}, we compare these constraints with those on modified dispersion relations from the propagation speed of GWs, while in section \ref{compa} we find a solar-system bound which is compared with the one from GWs. Section \ref{stqg} applies some of these results to specific theories, in particular, loop quantum cosmology, where quantum corrections are reinterpreted as a measurement of the difference between Hausdorff and spectral dimension in a near-infrared regime. Future developments in other models such as non-commutative inflation and varying-speed-of-light scenarios are briefly discussed in section \ref{oth}. The possibility to constrain quantum-gravity corrections with the instrumental noise of GW interferometers is presented in section \ref{stn}, where we give a prospect for LIGO, Virgo, KAGRA, LISA and DECIGO. Conclusions, where we also comment on future bounds from pulsars, are in section \ref{conc}. These results were in part anticipated in a short article \cite{QGld1}.


\section{Luminosity distance and GW amplitude in general relativity}\label{gr}

In this section, we review the definition of luminosity distance $d_L$ and the derivation of the dependence of the gravitational-wave amplitude on $d_L$ in GR, in a spacetime with $D$ topological dimensions (the physical case is $D=4$) and signature $(-,+,\cdots\!,+)$. We carry on this review  because it will point out, in a relatively familiar setting, the general scheme in which we will derive the expressions in the case of quantum gravity.

As a background, we take the flat homogeneous Friedmann--Lema\^itre--Robertson--Walker (FLRW) cosmological background with line element
\be\label{flrw}
\rmd s^2=-\rmd t^2+a^2(t)\,\rmd x_i \rmd x^i=a^2(\tau)\,(-\rmd \tau^2+\rmd x_i \rmd x^i)\,,
\ee
where $t$ is proper time and $\tau:=\int\rmd t/a$ is conformal time. We set the light speed $c=1$.


\subsection{Luminosity distance in GR}\label{gr1}

For a source emitting both gravitational waves and photons (electromagnetic source, EM), we can define two luminosity distances $d_L^\textsc{gw}$ and $d_L^\textsc{em}$ determined by observations.

By definition, the luminosity distance $d_L^\textsc{em}$ of an object emitting electromagnetic radiation is such that the flux ${\rm F}$ of light  reaching an observer is the power L per unit area emitted by the source, detected at a point on a sphere of radius $d_L^\textsc{em}$:
\be\label{ludi}
{\rm F}=:\frac{{\rm L}}{4\pi (d_L^\textsc{em})^2}\,.
\ee
In standard GR, the proper distance of a source emitting a single photon is $a(t_0)\,r=:a_0r$ measured by an observer at Earth at the present time $t_0$. Taking into account the redshift of power ${\rm L}=\textrm{(energy)}/\textrm{(time)}\propto (a/a_0)/(a_0/a)=a^2/a_0^2$ of photons reaching the observer at different times, one gets \cite{CoLu}
\be\label{ludi2}
d_L^\textsc{em} =\frac{a_0^2}{a}\,r\,.
\ee
In the absence of spatial curvature, $r$ can be written as $r=\tau_0-\tau(z)$, in terms of the redshift $1+z=a_0/a$. Setting $a_0=1$,
\be\label{dLGR}
d_L^\textsc{em}(z)=(1+z)\int^{t_0}_{t(z)}\frac{\rmd t}{a}=(1+z)\int^{1}_{a(z)}\frac{\rmd a}{H_\textsc{gr}a^2}=(1+z)\int_0^z\frac{\rmd z}{H_\textsc{gr}}\,,
\ee
where $a(z)=(1+z)^{-1}$. The Hubble parameter $H_\textsc{gr}(z)$ is determined by the first Friedmann equation and contains a parametrization of the dark-energy equation of state in terms of the barotropic index, for instance, $w=w_0={\rm const}$ or $w=w_0+(1-a)w_a$.

Expanding $H(z)$ for small $z$ and keeping only the lowest order, \Eq{dLGR} becomes
\be\label{dz1}
d_L^\textsc{em}\simeq\frac{z(1+z)}{H_0} \stackrel{z\ll 1}{\simeq} \frac{z}{H_0}\,,
\ee
where $H_0$ is the Hubble parameter today.


\subsection{Gravitational-wave amplitude in GR}\label{gwa}

The action and equations of motion in GR are
\be\label{eh}
S=\frac{1}{2\k^2}\int\rmd^Dx\,(R-2\Lambda)+S_{\rm matter}\,,\qquad R_{\mu\nu}-\frac12g_{\mu\nu}R+\Lambda g_{\mu\nu}=\k^2T_{\mu\nu}\,,
\ee
where $\k^2=8\pi G$ is Newton's constant, $\Lambda$ is a cosmological constant term (which we will ignore from now on), and $T_{\mu\nu}$ is the matter energy-momentum tensor. First we recall the expression of the GW amplitude in the local wave zone, and then consider its modification when the wave propagates on a homogeneous FLRW cosmological background.

\subsubsection{Local wave zone}\label{gwalw}

Let $h_{\mu\nu}$ be a metric perturbation around the Minkowski background $\eta_{\mu\nu}={\rm diag}(-,+,\cdots,+)$ and call $h$ one of the graviton polarization modes. The scalar $h$ is the amplitude of a gravitational wave emitted by a source such as a black-hole or a neutron-star binary system. We can express $h$ in terms of the luminosity distance $d_L^\textsc{gw}$, in $D$ topological dimensions. Expanding the Einstein equations to linear order in $g_{\mu\nu}=\eta_{\mu\nu}+h_{\mu\nu}$, one finds $\B_\eta h_{\mu\nu}=-2\k^2 S_{\mu\nu}$, where $\B_\eta=\eta^{\mu\nu}\p_\mu\p_\nu$ and $S_{\mu\nu}=T_{\mu\nu}-\eta_{\mu\nu} T_\rho^{\ \rho}/(D-2)$. The general retarded solution is given by the sum of the homogeneous solution (which will be ignored from now on) and the convolution of the source $S_{\mu\nu}$ with the retarded Green function associated with the kinetic operator $\B_\eta$ \cite{CDL}:
\ba
&&h_{\mu\nu}(x)=-2\k^2\int\rmd^Dx'\,S_{\mu\nu}(x')\,G^{\rm ret}(x-x')\,,\label{sou}\\
&&\B_\eta G^{\rm ret}(x-x')=\de^D(x-x')\,,\qquad G^{\rm ret}\big|_{t<t'}=0\,,\label{Gret}
\ea
where $\de^D(x-x')$ is the $D$-dimensional Dirac distribution. The retarded Green function in any dimension is \cite[eq.\ (22.37)]{Has13}
\be\label{ret1}
G^{\rm ret}(t,r)\propto \int\rmd\om\,\rme^{-\rmi\om t}\left(\frac{\om}{r}\right)^{\frac{D-3}{2}}H^{(1)}_{\frac{D-3}{2}}(\om r)\,,
\ee
where $r=|{\bf x}-{\bf x}'|$ and $H^{(1)}$ is the Hankel function of the first kind.

Since we are interested in the large-distance propagation of gravitational waves, we can approximate the above formul\ae\ in the \emph{local wave zone}, a region of space much larger than the size $s$ of the source and than the wave-length $\la\sim 1/\om$ of the waves. In this region,
\be\label{lwfa}
r\gg\la\sim s
\ee
and one can take the leading term of the $\om r\gg 1$ expansion of \Eq{ret1}. Since $H^{(1)}_{(D-3)/{2}}(\om r)\sim (\om r)^{-1/2}\rme^{\rmi\om r}$ up to a constant prefactor, one gets to leading order \cite{CDL}
\be\label{retsim}
G^{\rm ret}(t,r)\simeq c_G\frac{\de^{\left(\frac{D-4}{2}\right)}(t-r)}{r^{\frac{D-2}{2}}}\qquad\Rightarrow\qquad h_{\mu\nu}(t,r)\simeq \k \frac{c^h_{\mu\nu}(t-r)}{r^{\frac{D-2}{2}}}\,,
\ee
where $c_G$ is a constant that depends on the topological spacetime dimension $D$ (for $D=4$, $c_G=1/(4\pi)$), $\de^{\left(\frac{D-4}{2}\right)}$ is the derivative of order $(D-4)/2$ of the one-dimensional delta and $c^h_{\mu\nu}$ is a dimensionless tensorial function of retarded time.

Up to here, we have briefly reviewed the calculation of \cite{CDL}, omitting the details. Now we wish to rederive \Eq{retsim} by a purely dimensional argument, which will be sufficient to extract the $r$-dependence of the wave. In turn, the $r$-dependence will be what we need to obtain the luminosity-distance dependence of the GW amplitude on a cosmological background. This ``dimensionology'' will be extremely useful in the case of quantum gravity because it will spare us a full calculation of the GW amplitude. 

Consider a free scalar field on Minkowski spacetime, with action
\be
S_\vp=\int\rmd^D x\,\left(\frac12\vp \B_\eta\vp+\vp J\right),
\ee
where $J$ is a source. In terms of energy-momentum units, the volume density has engineering (energy) dimensionality $[\rmd^D x]=-D$, while the kinetic term has dimensionality $[\B]=2$. Since the action is dimensionless, the scalar has dimensionality
\be\label{vpD}
[\vp]=\frac{D-2}{2}\,.
\ee
Its retarded Green function obeys the distributional equation \Eq{Gret}, which tells us that $[G^{\rm ret}]=D-2$ in position space. This is consistent with \Eq{vpD}, since $G(x-x')=\langle\vp(x)\vp(x')\rangle$ is the two-point function of $\vp$. The inhomogeneous retarded solution of the equation of motion
\be\label{vJ}
\B_\eta\vp=(-\p_t^2+\N^2)\vp=-J
\ee
can be found as above by a convolution, $\vp (x)\,=\,-\int\rmd^Dx'\,J(x')\,G^{\rm ret}(x-x')$, but the dimensionality \Eq{vpD} of the scalar field already tells us that, to leading order in an $1/r$ expansion, the solution will scale as $\vp\sim r^{(2-D)/2}$ asymptotically far from the source. To see this, ignore the source $J$ in \Eq{vJ} and consider the radial profile $\vp(t,r)=r^{-(D-2)/2}\cF(t,r)$, where $\cF$ is a dimensionless function. In spherically-symmetric coordinates,
\ba
0&\simeq& (-\p_t^2+\N^2)\vp=-\p_t^2\vp+\frac{1}{r^{D-2}}\p_r(r^{D-2}\p_r\vp)\nonumber\\
&=&\frac{1}{r^\frac{D-2}{2}}\left[-\p_t^2+\p_r^2-\frac{(D-4)(D-2)}{4r^2}\right]\cF\,.\label{flat0}
\ea
In $D=4$ dimensions, the general solution is any function of retarded and advanced time, $\cF=\cF_1(t-r)+\cF_2(t+r)$ (compare with \Eq{retsim}). For $D\neq 4$, we can take this as an approximate solution as long as the wave frequency is sufficiently large, $\om r\gg\sqrt{(D-4)(D-2)/4}$. This condition always holds in the wave zone. Therefore, taking only the retarded function one has
\be
\vp\simeq \frac{\cF(t-r)}{r^\frac{D-2}{2}}\,.
\ee

Going back to the GR action \Eq{eh}, the Newton's constant has dimension $[\k^2]=2-D$, negative for $D>2$, while the gravitational field is dimensionless, $[h_{\mu\nu}]=0$. From now on, we consider the amplitude $h$ and drop tensorial indices. The rescaled field $\bar h:=h/\k$ has the dimensionality of an ordinary scalar field, $[\bar h]=(D-2)/2$, and the above argument for $\vp$ applies almost \emph{verbatim} to $\bar h$. Thus, in the local wave zone
\be\label{lwz}
h\simeq \frac{\k\cF_h(t-r)}{r^{\frac{D-2}{2}}}\stackrel{D=4}{\propto}\frac{1}{r}\,,
\ee
where $\cF_h$ is a function of retarded time.

\subsubsection{Cosmological propagation}

The amplitude \Eq{lwz} is what one would detect in the local wave zone, a neighborhood of the astrophysical source much larger than the wavelength of the signal,  but small enough to be insensitive to the cosmic expansion. If the observer is at a cosmological distance, the expansion must be taken into account.

This step is relatively easy once the local radial dependence of the wave form has been found. In fact, the luminosity distance \Eq{ludi2} is proportional to $r$ and $r$ in \Eq{lwz} is nothing but the comoving distance from the source. This suggests that, up to various rescalings with the scale factor $a$ and up to a function dependent on emission and observation time,
\be\label{hcos}
h\sim \frac{1}{(d_L^\textsc{em})^{\frac{D-2}{2}}}\stackrel{D=4}{=}\frac{1}{d_L^\textsc{em}}\,.
\ee
This is indeed correct, as one can carefully check in four dimensions \cite[section 4.1.4]{Mag07}. Again, a dimensional argument (a generalization to $D$ dimensions of an observation made in \cite{Mag07} for $D=4$) can help to understand the final result \Eq{hcos}. Consider our scalar field $\vp$ with source $J$, this time propagating on a flat FLRW background with line element \Eq{flrw}. On this background, the equation of motion is
\be
\B\vp=-\frac{1}{a^2}[\p_\tau^2+(D-2)\cH\p_\tau-\N^2]\vp=-J\,,\qquad \cH:=\frac{\p_\tau a}{a}\,.
\ee
The engineering dimensionality \Eq{vpD} indicates that a natural rescaling of the field is
\be\label{wgr}
w:=a^{[\vp]}\vp=a^{\frac{D-2}{2}}\vp\qquad\Rightarrow\qquad \left(\p_\tau^2-\frac{\p_\tau^2 a}{a}-\N^2\right)w=a^{\frac{D+2}{2}}J=:\cJ\,,
\ee
where $\N^2$ is the Laplacian in comoving coordinates ${\bf x}$ and $\cJ(\tau,{\bf x})$ is an effective source term. In a matter-dominated universe, $a\sim\tau^2$, so that $\p_\tau^2 a/a\simeq 2/\tau^2$, which is negligible with respect to the other terms inside the Hubble horizon. Dropping this time-dependent effective mass, we are left with the conformally flat equation $(-\p_\tau^2+\N^2)w=-\cJ$. Therefore, starting from here one can use the results of the previous subsection in Minkowski spacetime, with time $t$ replaced by conformal time $\tau$ and $\vp$ (or $\bar h$) replaced by $w$. In particular, $w\simeq r^{-(D-2)/2}F_w(\tau-r)$ and $\vp\sim (ar)^{-\frac{D-2}{2}}$. Applying this $r\to ar$ rule to \Eq{lwz} and evaluating it today, $h\sim (a_0r)^{-\frac{D-2}{2}}=[(1+z)/d_L^\textsc{em}]^{\frac{D-2}{2}}$. The redshift factor is absorbed into the source function we have not calculated, and one obtains \Eq{hcos} as announced.


\section{Beyond GR: parametrizing luminosity distance and Planck mass}\label{para}

Information from gravitational waves can be used in such a way as to construct observables related to how spacetime affects GW propagation. In general, theories beyond GR can change both the dispersion relation of GWs and the background metric solutions, or other aspects of the dynamics in various ways. Independently of their details, we can encode these effects into the ratio between the GW and the electromagnetic luminosity distance at any redshift,
\be\label{dez3}
\Xi(z):=\frac{d_L^\textsc{gw}(z)}{d_L^\textsc{em}(z)}\,.
\ee

In certain modified gravities \cite{Lin18,BDFM}, the parameter \Eq{dez3} is related to the effective Newton constant measured in structure-formation observations. Consider the ordinary momentum-space Poisson equation $-k^2\Phi_k=a^2\rho_{\rm m}\De^{\rm m}_k/(2M_\Pl^2)$, where $\Phi$ is the Newton potential (scalar perturbation of the 00-component of the metric, generated by matter during structure formation), $\rho_{\rm m}$ is the background matter density and $\De^{\rm m}_k$ is the comoving density perturbation in momentum space. In models where gravity is modified, one can write down a generic effective Poisson equation for $\Phi$, with effective momentum  and time-dependent Planck mass (Newton constant) $M_{\rm eff}$ \cite{AKS,Dan10}:
\be\label{pois}
-k^2\Phi_k=\frac{a^2\rho_{\rm m}\De^{\rm m}_k}{2M_{\rm eff}^2(t,k)}\,.
\ee
This is the definition of $M_{\rm eff}^2$ in $D=4$ dimensions. We will not use this parameter much in the following, since the profile $M_{\rm eff}^2(t,k)$ in \Eq{pois} is model dependent and we rather focus on general effects. In typical classical modified-gravity models where spacetime geometry possesses no fundamental scale, $M_{\rm eff}$ is purely homogeneous \cite{BDFM}.

In GR, gravitational and electromagnetic waves have the same dispersion relation: they both propagate with the speed of light, and the luminosity distance defined by \Eq{ludi} is equivalent to $d_L^\textsc{gw}$ at all scales. Therefore, $\Xi(z)=1$. When the dynamics or the very structure of spacetime are modified, this equivalence no longer holds, and one must parametrize the right-hand side of \Eq{dez3}  as a function of $z$. In the next section, we will see how these modifications arise and what parametrizations can be employed. Here we will use the parametrization of \cite{BDFM}:
\be\label{eq:param}
\Xi = \Xi_0 + (1-\Xi_0) a^{n}=\Xi_0 + \frac{1-\Xi_0}{(1+z)^n}\stackrel{z\ll 1}{=} 1-(1-\Xi_0)nz+O(z^2) \,,
\ee
where $\Xi_0>0$ and $n$ are constant. In GR, $\Xi_0=1$.

From the $\k$-dependence of the metric perturbation in \Eq{retsim}, it is not difficult to convince oneself that $d_L^\textsc{gw}$ is proportional to a power of the effective Planck mass appearing in front of the graviton kinetic term when expanding the action to $O(h^2)$ around a FLRW background,
\be\label{hboxh}
\de S=\frac12\int\rmd^Dx\,\sqrt{-g}\,M^{D-2}_\textsc{gw}\,h_{ij}\B_\textsc{flrw} h^{ij}\,.
\ee
In GR, $M^{2-D}_\textsc{gw}=\k^2=8\pi G=M_\Pl^{2-D}$ is the reduced Planck mass, but in other theories (starting with purely classical scalar-tensor and $f(R)$ models) it happens that $M_\textsc{gw}\neq M_\Pl$ and that $M_\textsc{gw}(t)$ is a function of the cosmological background and acquires a non-trivial time dependence. By the same token, we can recast the parameter $\Xi$ expressing the ratio of luminosities as a function of the effective mass $M_{\rm eff}$. Thus, in some classical models
\be\label{dM}
\Xi(z)=\left[\frac{M_{\rm eff}(z)}{M_\textsc{gw}(z)}\right]^{\frac{D-2}{2}}.
\ee
This expression is valid only under the assumption that all the corrections to the dynamics can be encoded at
sufficiently large scales in an effective Newton constant. In general, this happens when the comoving number density of gravitons is conserved \cite{BDFM}. However, there are cases where \Eq{dM} may not hold, as in higher-dimensional braneworld models or in models where the graviton is unstable \cite{BDFM}. We will see later that also in quantum gravity the left- and right-hand sides of \Eq{dM} may be different quantities, the reason being that volumes (including also comoving volumes) do not scale as expected in certain regimes. Therefore, if one is interested in placing constraints on effective Planck masses, it is useful to consider also other parameters aside from ratios of luminosity distances.

A parameter first introduced in  dark-energy models based on modified gravity is the time variation of the effective Planck mass $M_\textsc{gw}$ \cite{BeSa,GLV}, which we extend here to $D$ dimensions,
\be\label{aM} 
\a_M:=\frac{\rmd\ln M^{D-2}_\textsc{gw}}{\rmd\ln a}=-\frac{\rmd\ln M^{D-2}_\textsc{gw}}{\rmd\ln (1+z)}\,,
\ee
and we will further modify to include quantum-gravity effects.\footnote{This parameter is called $\nu$ in \cite{Nis17} and $-2\de$ in \cite{BDFM}.} Its origin is the action \Eq{hboxh}. Contracting out the polarization matrix and absorbing the mass factor into the dimensionful field $\bar h=M^{(D-2)/2}_\textsc{gw} h$, \Eq{hboxh} is recast as $\de S=(1/2)\int\rmd^Dx\,\sqrt{-g}\,\bar h\B_\textsc{flrw}\bar h+\dots$, yielding the equation of motion
\be
0=\bar h''+(D-2)\cH\bar h'+O(\bar h)=M^{\frac{D-2}{2}}_\textsc{gw}\left\{h''+[(D-2)\cH+\a_M] h'+O(h)\right\},
\ee
where primes are derivatives with respect to conformal time.

Yet another effective mass of interest appears in the first Friedmann equation,
\be\label{frw1}
H^2=\frac{\rho(t)}{3M_\textsc{flrw}^{D-2}(t)}+\dots\,,
\ee
where $\rho$ is the energy density of the content of the universe and ``$\dots$'' are other corrections to the GR dynamics. In general, \Eq{frw1} is not sufficient to describe the cosmological effects of modified gravity, since, on one hand, also the background continuity equation may receive corrections and, on the other hand, inhomogeneities are also affected, with consequences for cosmological perturbations and structure formation.

In GR, $M_\textsc{flrw}=M_{\rm eff}=M_\textsc{gw}=M_\Pl$. However, in models with screening effects, such as the chameleon mechanism \cite{KhWe1,KhWe2}, the effective Planck mass measured in local ($z\ll 1$) observations may differ from the effective Planck mass seen by gravitational waves propagating at cosmological distances. Therefore, as last parameters we define the relative difference between $M_\textsc{flrw}$ (or $M_{\rm eff}$) and $M_\textsc{gw}$:
\be\label{MeffM}
\De M_\textsc{flrw}:=\left(\frac{M_\textsc{flrw}}{M_\textsc{gw}}\right)^{D-2}-1\,,\qquad \De M_{\rm eff}:=\left(\frac{M_{\rm eff}}{M_\textsc{gw}}\right)^{D-2}-1\,.
\ee


\section{Dimensional flow in quantum gravity}\label{qgs}

To the best of our knowledge, all theories of quantum gravity display dimensional flow,\footnote{Also known as ``dimensional reduction,'' a term often used in the literature that may create confusion with the compactification schemes of supergravity and string theory.} a phenomenon according to which at least one or more of the dimensions of spacetime change with the probed scale \cite{tH93,Car09,fra1,revmu,Car17}. There are several inequivalent definitions of dimension, the most used ones being:
\begin{itemize}
\item the topological dimension $D$ (the number of spacetime coordinates, which is fixed and constant at the outset);
\item the Hausdorff dimension $\dh$ of spacetime (governed by the position-space measure);
\item the Hausdorff dimension $\dh^k$ of momentum space;
\item the spectral dimension $\ds$ (governed by the type of kinetic term in the action; it depends on the dispersion relation and on the measure in momentum spacetime).
\end{itemize}
In this section, we review these quantities in general and recall or calculate their value for specific theories. Then we will extract a scaling parameter that will enter the GW amplitude. Conventionally, all dimensions are defined in Euclideanized spacetime.

To describe dimensional flow, let us simplify the problem to a free scalar field on flat spacetime with action
\be\label{Svqg}
S=\frac12\int\rmd^D x\,v(x)\,\vp \cK\vp\,,
\ee
where $v$ is the spacetime measure weight and $\cK$ is a kinetic term. The action $S$ is imagined to stem from a theory of quantum gravity in the continuum limit (if continuous spacetime is not a fundamental structure of the theory, as in all approaches with discreteness and non-commutativity) and in a local inertial frame. The latter point is important: ``quantum gravity'' does not necessarily mean ``curvature corrections'' and one can have non-trivial effects even in low-curvature regimes. We encode these effects into a non-trivial weight $v\neq 1$ and a generalized kinetic term $\cK\neq\B$. If the theory is Lorentz-invariant in the continuum limit, then $\cK=\cK(\B)$, but for our discussion it can be more general (for instance, a fractional operator \cite{revmu}).

The theories of QG we will consider in this paper will be GFT/spin foams/LQG, causal dynamical triangulations, $\k$-Minkowski spacetime, Stelle gravity, the low-energy limit of string theory, asymptotic safety, Ho\v{r}ava--Lifshitz gravity, and a non-local black-hole model by Padmanabhan. This list, encompassing theories where the Hausdorff and spectral dimension have been computed, does not exhaust the literature. For instance, we do not discuss causal sets, nor other theories for which dimensional flow has not been studied. Also, little is known about the cosmology of some of these theories, as in the case of causal dynamical triangulations, $\k$-Minkowski spacetime\footnote{However, we will see a non-commutative cosmological model in section \ref{ncc}.} and Padmanabhan's model. In all cases, however, our findings from GWs in sections \ref{ludis} and \ref{nums} will be the first of their kind.

For simplicity, we will employ definitions of dimensions (\Eq{dh}, \Eq{dhk} and \Eq{ds}) valid on continuous spacetimes, but they can be generalized to arbitrary sets of points. Several theories of quantum gravity admit a notion of continuous spacetime, but for GFT/spin foams/LQG discreteness effects dominate below a certain scale (close to the Planck scale) where the underlying simplicial-complex structure becomes manifest. In this case, eqs.\ \Eq{dh}, \Eq{dhk} and \Eq{ds} must be replaced with other definitions we will not discuss here \cite{COT3}, valid at arbitrarily small scales.

Regarding nomenclature, we will use UV and IR labels to characterize regimes where corrections to general relativity are, respectively, important or negligible. Thus, independently of whether they are Planckian, atomic, solar-system size or cosmological:
\begin{itemize}
\item scales where general relativity is fully recovered will belong to the \emph{infrared} (IR) or \emph{general-relativistic} (GR) regime of dimensional flow;
\item intermediate scales where the corrections to GR are small but non-negligible will be called \emph{mesoscopic} (meso). Mesoscopic scales particularly close to the IR regime will be occasionally called \emph{near-IR}; 
\item scales at which quantum-gravity corrections are important will belong to an \emph{ultraviolet} (UV) or \emph{quantum-gravity} (QG) regime.
\end{itemize}
 As far as the running of spacetime dimensions is concerned, the IR and the UV regimes correspond, respectively, to large and short scales, while mesoscopic scales can be near-Planckian or very large. However, the phenomenology of dimensional flow is more complex than dimensional flow itself, and there exist cosmological QG models where this correspondence between the IR and large scales on one hand, and the UV and short scales on the other hand, is not reflected on physical observables. Therefore, a model where GR is recovered at cosmological scales and QG corrections dominate at short scales will be characterized by an IR or mesoscopic large-scale regime and a UV short-scale regime; here, QG should be tested at atomic or particle-physics scales. Conversely, a model where QG corrections dominate at cosmic scales and die away locally will be characterized by a UV large-scale regime and an IR or mesoscopic short-scale regime, in which case astrophysical and cosmological observations may be more interesting. 

To prevent confusion in what follows, let us reiterate the point: whenever we talk about UV and IR scales, we will refer to a QG and to a GR regime, respectively, while the relative size of the scales at which these regimes take place will be denoted by adjectives such as ``large,'' ``cosmological,'' ``small,'' ``local,'' and so on. By definition, mesoscopic scales are intermediate both in terms of QG effect and in terms of relative size.


\subsection{Hausdorff dimension of spacetime}

The Hausdorff dimension $\dh$ is the scaling of the volume $\cV$ of a $D$-ball (or a $D$-cube) with respect to the radius (edge size) $\ell$. Assuming a continuum ambient spacetime, the definition is
\be\label{dh}
\dh(\ell):=\frac{\rmd\ln\cV(\ell)}{\rmd\ln\ell}\,.
\ee
For a space or spacetime without dimensional flow, $\cV=\int_{\rm cube}\rmd^D x\,v(x)\propto\ell^{\dh}$ and $\dh$ is constant. In ordinary spacetime, or in the IR limit of spacetimes with dimensional flow, $\dh=D$. For a fractal, $\dh$ can be non-integer. Although many quantum theories of gravity do not admit spacetimes with a varying Hausdorff dimension, all of them admit spacetimes with varying spectral dimension. 

\begin{figure}
\centering
\includegraphics[width=10cm]{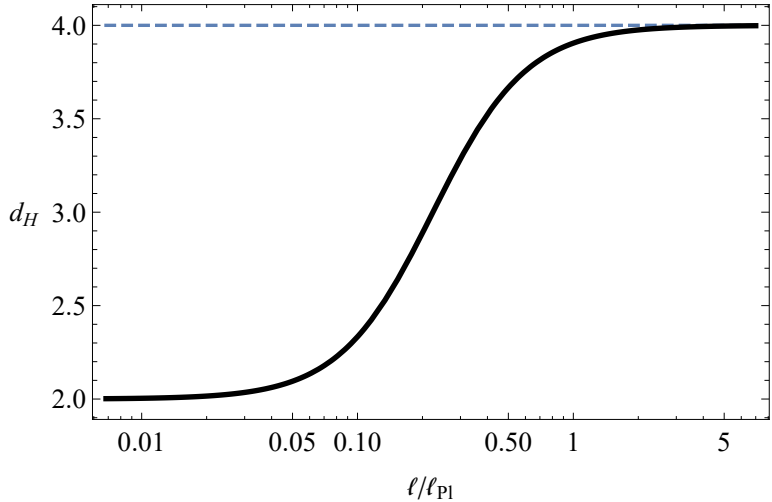}
\caption{\label{fig1} Representative profile for the spacetime Hausdorff dimension of quantum states of geometry in GFT/spin foams/LQG (solid thick curve), as a function of the length scale ratio $\ell/\lp$. Here $\ell_*=\lp$ is the Planck scale. This profile is obtained from an \emph{ad hoc} analytic expression of the type $\dh(\ell)=1+[3(\ell/\ell_*)^2+c]/[(\ell/\ell_*)^2 +c]$, following the discussion in \cite{frc4}. It qualitatively reproduces the actual profile calculated numerically in \cite{COT3}. The dashed line is the IR asymptote, i.e., the Hausdorff dimension of theories with constant $\dh=D=4$.}
\end{figure}

String field theory, asymptotic safety, causal dynamical triangulation, non-local quantum gravity, most formulations based on non-commutative spacetimes, Ho\v{r}ava--Lifshitz gravity and a few others, all have constant Hausdorff dimension, which coincides with the topological dimension $\dh=D$. In these theories, the main effect is a modification of the effective dispersion relation and, hence, a running spectral dimension (to
be discussed below). However, there are theories where not only $\ds$, but also $\dh$ varies, as for instance in the case of cyclic-invariant non-commutative theories on $\k$-Minkowski spacetime \cite{AAAD}, the group of theories based on discrete pre-geometries (GFT/spin foams/LQG) and multi-fractional theories. In cyclic-invariant models and in GFT/spin foams/LQG, the Hausdorff and spectral dimension of spatial slices vary \cite{ACOS,COT2,COT3,MiTr}, which means that a perfectly homogeneous cosmology is insensitive to this specific feature. However, multi-fractional theories admit a varying $\dh$ also along the time direction. Since GFT/spin foams/LQG will draw special attention later, in Fig.\ \ref{fig1} we show a representative profile for the spacetime Hausdorff dimension as function of the scale. Note that this profile is valid also at ultra-short scales where discreteness effects are important and \Eq{dh} does not hold.


\subsection{Hausdorff dimension of momentum space}

Given a continuous momentum space with measure weight $w(k)$, its Hausdorff dimension is the scaling of the volume $\tilde\cV(k)=\int_{\rm cube}\rmd^Dk'\,w(k')$ of a $D$-cube (or $D$-ball) of linear size (respectively, radius) $k$:
\be\label{dhk}
\dh^k(\ell):=\frac{\rmd\ln\tilde\cV(k)}{\rmd\ln k}\Big|_{k=1/\ell}\,.
\ee
$\k$-Minkowski spacetime is the only example we will consider in this paper where there is dimensional flow in $\dh^k$. We can calculate $\dh^k$ from the Euclidean momentum-space measure $\rmd^4p/\sqrt{p^2+\k^2}$, where $p=p(k)$ is a redefinition of momenta. Integrating it, one obtains the momentum 4-volume $\cV(p)\propto (p^2-2\k^2)\sqrt{p^2+\k^2}+2\k^4$, where $\k^2=8\pi/\lp^2$ and the constant term $2\k^4$ has been added to recover the ordinary volume $\cV(p)\sim p^4+O(p^6)$ in the large-scale limit $p\to 0$. Thus,
\be\label{dhkmink}
\dh^k(\ell)=\frac{\rmd\ln\cV(p)}{\rmd\ln p}\Big|_{p=1/\ell}=\frac{3}{1-(\k\ell)^2+2(\k\ell)^4[\sqrt{1+(\k\ell)^{-2}}-1]}\,.
\ee
Then, $\dh^k\simeq 3$ at small scales $\k\ell\ll 1$, while $\dh^k\simeq 4$ at large scales $\k\ell\gg 1$.

All the other QGs discussed here have constant momentum-space Hausdorff dimension, equal to the topological dimension $\dh^k=D$. In particular, in GFT/spin foams/LQG the momentum space is compact but it has the ordinary measure below the compactness cut-off (again, where discreteness dominates) \cite{COT3,MiTr}, so that $\dh^k=4$ in the continuum regime we are considering.


\subsection{Dispersion relations and spectral dimension}\label{dires}

In general, in a spacetime with dimensional flow the scaling $[\vp]$ depends on the scale $\ell$ one is observing, and $[\vp]_{\rm UV}\neq [\vp]_{\rm IR}$. Consider for illustrative purposes (and ignoring unitarity issues) the Lorentz-invariant form factor $\cK(\B)=\ell_*^{2-2\b}\B+(-\B)^\b$, corresponding to the dispersion relation
\be\label{dr1}
\cK(-k^2)=-\ell_*^{2-2\b}k^2+k^{2\b},\qquad [\cK]=2\b\,,
\ee
where $\b>1$ and momentum-space coordinates have the usual dimensional units $[k^\mu]=1$. Such form factor or similar ones arise in higher-order non-unitary theories ($\b=2,3,\dots$) \cite{Ste77,ALS,AAM}, the propagation of low-energy particles in non-critical string theory ($\b=3/2$ at mesoscopic scales and $\b=2$ in the deep UV) \cite{ACEMN}, asymptotic safety ($\b=2$) \cite{LaR5}, Ho\v{r}ava--Lifshitz gravity ($\b=3$ only in the spatial directions) \cite{Hor3}, multi-fractional spacetimes with fractional and $q$-derivatives ($\b>0$ real; Lorentz invariance is broken here) \cite{frc4,frc7}, causal sets ($\b=2$) \cite{BBL},  and as an effective dispersion relation in loop quantum gravity \cite{GaPu,AMTU,ACAP,Ron16}. Other cases such as non-local quantum gravity do not correspond to a polynomial dispersion relation such as \Eq{dr1} and must be treated separately \cite{BrCM}. Last, $\ell_*$ is a length scale around which quantum-gravity effects become relevant. If there is only one fundamental scale $\ell_*$, then it is very small (of order of, or not too far from, the Planck scale $\lp$) and the UV regime corresponds to very short wave-lengths. If, however, \Eq{dr1} is an effective dispersion relation, $\ell_*$ is a mesoscopic scale that could be relatively far from the actual UV and closer to the IR regime. In order to keep the following discussion as flexible as possible, and in line with the nomenclature declared at the beginning of this section, it will be understood that the subscript or superscripts ``UV,'' ``meso'' and ``IR'' indicate scalings and dimensions at scales where QG corrections are, respectively, large, small and completely negligible, independently of the actual size of these scales. Thus, a UV scale does not necessarily correspond to the microscopic end of the dimensional running. To cover both the UV and mesoscopic regime at the same time, we will use a non-committal subscript or superscript * as in $\ell_*$.

Let us recall the definition of the spectral dimension $\ds$ (see \cite{CMNa} for more details and references). Consider the dispersion relation $\tilde\cK(-k^2)=\ell_*^{2\b-2}\cK(-k^2)$ with dimensionality $({\rm energy})^2$ and the heat kernel
\be\label{P}
P(x-x';\ell)=\frac{1}{(2\pi)^D}\int\rmd^Dk\,w(k)\,\rme^{\rmi k\cdot (x-x')}\rme^{-\ell^2\tilde\cK(-k^2)}\,,
\ee
where $w(k)$ is a measure weight that accounts for quantum-gravity effects.\footnote{In general, the eigenfunctions of the operator $\cK$ may carry a scale dependence which combines with the actual measure in momentum space. The weight $w(k)$ is the outcome of this combination.} For instance, momentum spaces with non-trivial measures are typically associated with non-commutative spacetimes (e.g., \cite{KoWa,ACo} and references therein). In Schwinger representation, the Green function of the scalar field theory \Eq{Svqg} is
\be\label{scw}
G(x-x')=\ell_*^{2\b-2}\int_0^{+\infty}\rmd\ell^2\,P(x-x';\ell)=-\frac{1}{(2\pi)^D}\int\rmd^Dk\,w(k)\,\rme^{\rmi k\cdot (x-x')}\frac{1}{\cK(-k^2)}\,.
\ee
From the heat kernel, one finds the so-called return probability
\be\nonumber
\cP(\ell)=P(0;\ell)=\frac{1}{(2\pi)^D}\int\rmd^Dk\,w(k)\,\rme^{-\ell^2\tilde\cK(-k^2)}\,,
\ee
from which one defines the spectral dimension
\be\label{ds}
\ds:=-\frac{\rmd\ln\cP(\ell)}{\rmd\ln\ell}\,.
\ee
For a spacetime without dimensional flow, $\cP\propto\ell^{-\ds}$ with $\ds$ a constant. In ordinary spacetime, or in the IR limit of spacetimes with dimensional flow, $\ds^{\rm IR}=D$. In the case of the dispersion relation \Eq{dr1}, the spectral dimension in the UV/meso regime is easy to compute. If the momentum measure in this regime goes as $\rmd k\,k^{\dh^{k,*}-1}$, and defining the dimensionless variable $y:=(\ell/\ell_*)^2 (\ell_* k)^{2\b}$, we find that $\cP=(\ell_*^{\b-1}\ell)^{-\dh^{k,*}/\b}\int_0^{+\infty}\rmd y\,F(y)$, where $F$ is a function of $y$, so that
\be\label{dsuv}
\ds^*=\frac{\dh^{k,*}}{\b}\,.
\ee
With a dispersion relation of the type \Eq{dr1}, with anomalous dimensionality (energy)$^{2\b}$, the scalar field $\vp$ has the energy dimensionality (remember that $[S]=0$)
\be\label{vpuv0}
[\vp]=[\vp]_*=\frac{D\a-2\b}{2}=\frac{\dh^*}{2}-\frac{\dh^{k,*}}{\ds^*}\,.
\ee
Consistently, the dimensionality of the Green function from \Eq{scw} is $[G]=\dh^{k,*}-2\b$. 

If we had chosen the dispersion relation $\cK=-k^2+\ell_*^{2\b-2}k^{2\b}$, with the $\ell_*$ factor attached to the UV term, then $[\cK]=2$, the heat kernel \Eq{P} would be the same but with $\tilde\cK(-k^2)=\ell_*^{2}\cK(-k^2)$, the Schwinger representation \Eq{scw} would have changed by the replacement $\ell_*^{2\b}\to\ell_*^2$, the Green function would have had dimensionality $[G]=D-2$, and the scalar field would have had ordinary dimensionality $[\vp]=[\vp]_{\rm IR}=(D-2)/2$. After a suitable change of coordinates $k\to\tilde k$ in momentum space \cite{AAGM3} or a redefinition of \Eq{ds}, one can directly get $\ds=\dh^{\tilde k}$. This is just a matter of convention, which does not affect the final result \Eq{dsuv}. Since the UV scaling is the one entering power-counting arguments on renormalization, we stick with the UV scalings discussed above. The dimensionality of $\vp$ in the IR stems from redefining the mode as $\vp_{\rm IR}=\ell_*^{[\vp]-[\vp]_{\rm IR}}\vp$, so that $[\vp]_{\rm IR}=[\vp_{\rm IR}]$. 

\begin{figure}
\centering
\includegraphics[width=10cm]{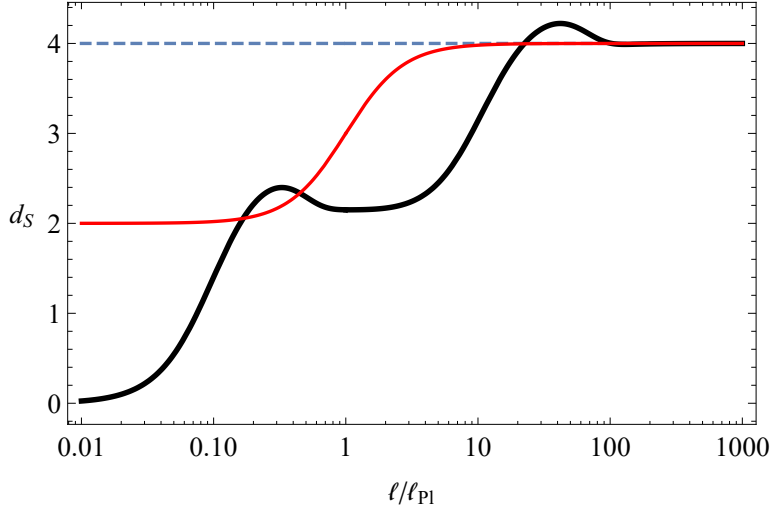}
\caption{\label{fig2} Typical spectral dimension of spacetimes in GFT/spin foams/LQG (solid thick black curve) and theories with monotonic flow from below with $\ds^{\rm UV}=2$ (solid thin red curve), where $\ell$ is a length scale. These profiles are obtained from \emph{ad hoc} analytic expressions following the discussion in \cite{frc4}, reproducing the actual profiles for these theories (calculated numerically in the case of GFT/spin foams/LQG \cite{COT3}). For GFT/spin foams/LQG, discreteness effects start to become important at $\ell/\lp\lesssim 1$. In this figure, the mesoscopic scale $\ell_*$ is around the local maximum at $\sim 40\,\lp$. The dashed line is the IR asymptote.}
\end{figure}

In Fig.\ \ref{fig2}, we show the spectral dimension of Stelle theory \cite{CMNa}, asymptotic safety \cite{LaR5} and Ho\v{r}ava--Lifshitz gravity \cite{Hor3}, where $\ds$ flows monotonically from $\ds^{\rm UV}=2$ up to 4. The spectral dimension of GFT/spin foams/LQG, also shown in the figure, was calculated numerically in \cite{COT3} for simplicial-complex quantum states of geometry, without using any continuum effective limit.


\subsection{A scaling parameter}

Overall, we can write all expressions in a scale-independent way. For instance, we call $\dh^{k}=[\rmd^Dk\,w(k)]$ the effective Hausdorff dimension of momentum spacetime regardless of the scale; in the most general case, $\dh^{k}\neq\dh$. Also, instead of \Eq{dsuv} we have
\be\label{dsgen}
\ds=2\frac{\dh^k}{[\cK]}\,.
\ee
The generalization of \Eq{vpuv0} to all scales takes the simple form $[\vp]=\Gamma$, where
\be\label{vpuv}
\boxd{\Gamma:=\frac{\dh-\tilde{d}_{\rm W}}{2}\,,\qquad \ds\neq 0\,,}
\ee
and we defined the effective walk dimension
\be
\tilde{d}_{\rm W}:=2\frac{\dh^k}{\ds}\,,
\ee
which is nothing but the dimensionality $[\cK]$ of the kinetic operator. On a geometry where $\dh^k=\dh$, this indicator coincides with the walk dimension \cite{bAH}, the scaling of the variance of a random walker in spacetime, $\langle X^2\rangle\propto \s^{2/{\dw}}$, where $\s$ is a fictitious diffusion time. We will reserve the symbol $\Gamma_{\rm UV}$ for the scaling \Eq{vpuv} in the UV and $\G_{\rm meso}$ for the scaling \Eq{vpuv} where QG corrections are small but non-negligible. We repeat that, depending on the model, UV scales do not necessarily correspond to the shortest (Planck-size) ones, nor mesoscopic scales are necessarily large (super-atomic).

\begin{figure}
\centering
\includegraphics[width=10cm]{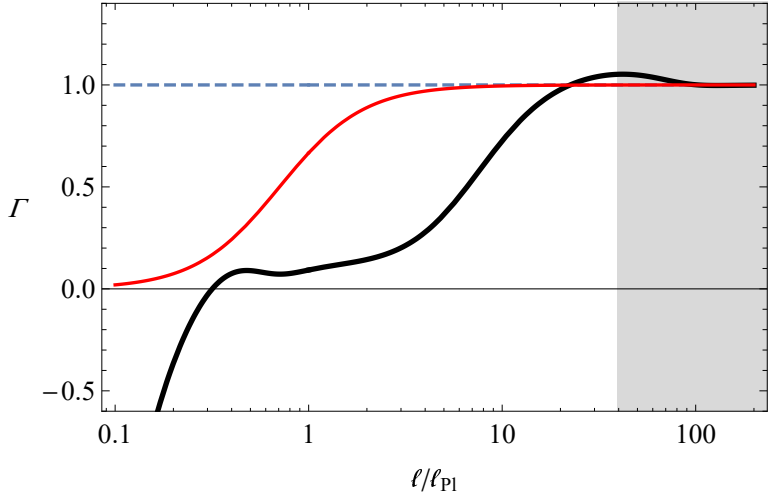}
\caption{\label{fig3} The scaling parameter \Eq{vpuv} in GFT/spin foams/LQG (solid thick black curve) and theories with monotonic flow from below with $\ds^{\rm UV}=2$ (solid thin red curve), where $\ell$ is a length scale. These curves are obtained from the \emph{ad hoc} analytic profiles shown in Figs.\ \ref{fig1} and \ref{fig2}. The region where discreteness effects become important in GFT/spin foams/LQG (black thick curve plunging down to $\G<-3$) is not shown in the plot. In this figure, the mesoscopic scale $\ell_*$ is around the local maximum at $\sim 40\,\lp$; this scale marks the border of the shaded region where $0<\G_{\rm meso}-1\ll1$. The dashed line is the IR asymptote.}
\end{figure}

In the IR, $\ds=\dh=D$ and one recovers \Eq{vpD}. For the special value $\ds^{\rm UV}=2$ (string theory, Stelle gravity, asymptotic safety and Ho\v{r}ava--Lifshitz gravity), the field is dimensionless when $\dh^k=\dh$, a fact which has consequences for the renormalization of the theory. In fact, \Eq{vpuv} shows that a spectral dimension $\ds^{\rm UV}=2$ drives the field to a regime where the field is dimensionless and, consequently, the interaction coupling constant in the action has positive dimensionality, which, in turn, guarantees power-counting renormalizability. Thus, with a simple scaling argument, we can understand one of the \emph{leitmotifs} in quantum gravity surrounding the ``magic'' number $\ds^{\rm UV}=2$ \cite{tH93,Car09}, although this argument only applies to QGs where the UV can be treated as a perturbative field theory.\footnote{In general, power-counting renormalizability can cease to have meaning in theories where the deep UV regime can be described only with non-perturbative techniques.} The spectral dimension and the parameter \Eq{vpuv} for these theories are shown in Fig.\ \ref{fig2} and \ref{fig3}, respectively; here $\dh^k=\dh=D$ at all scales, so that $\Gamma=(\ds-2)D/(2\ds)$. In theories where $\ds=\dh=\dh^k$ at all scales, $\Gamma=(\dh-2)/2=(\ds-2)/2$. Theories with $\ds^{\rm UV}=0$ (for instance, non-local quantum gravity) are a special case we deal with elsewhere \cite{BrCM}. The values of the UV spacetime dimensions and of the parameter $\Gamma_{\rm UV}$ in several quantum gravities are reported in Tab.\ \ref{tab1}.
\begin{table}
\centering
\begin{adjustwidth}{-.05cm}{}
\begin{tabular}{l|c|c|c|c|c}\hline
																														 &$\dh^{\rm UV}$&$\dh^{k,{\rm UV}}$& $\ds^{\rm UV}$ & $\Gamma_{\rm UV}$ & $\Gamma_{\rm meso}\gtrsim 1$\\\hline\hline
GFT/spin foams/LQG \cite{ACAP,COT3,MiTr} 										 & 2   					& 4								 &$[1,4)$&$[-3,0)$ & \ding{51} \\
Causal dynamical triangulations (phase C) \cite{CoJu}	       & 4						& 4 						   & $3/2$          		& $-2/3$ & \\
$\k$-Minkowski bicovariant $\N^2$ \cite{Ben08,ArTr} (c.i.m.) & 1   					& 3     					 & 3	      					& $-1/2$ & \\
$\k$-Minkowski bicross-product $\N^2$ \cite{ArTr} (c.i.m.)   & 1   					& 3     					 & 6      						&	0      & \\
Stelle gravity \cite{CMNa}    														   & 4            & 4                & 2              		& 0      & \\
String theory (low-energy limit) \cite{ACEMN,CaMo1}					 & $D$          & $D$              & 2              		& 0      & \\
Asymptotic safety \cite{LaR5}    														 & 4            & 4                & 2              		& 0      & \\
Ho\v{r}ava--Lifshitz gravity \cite{Hor3}										 & 4            & 4                & 2              		& 0      & \\
$\k$-Minkowski relative-locality $\N^2$ \cite{ArTr} (c.i.m.) & 1   					& 3     					 & $+\infty$  				& $1/2$  & \\
$\k$-Minkowski bicovariant $\N^2$ \cite{Ben08,ArTr} (o.m.)   & 4   					& 3     					 & 3	      					& 1 &  \\
$\k$-Minkowski bicross-product $\N^2$ \cite{ArTr} (o.m.)     & 4   					& 3     					 & 6      						&	$3/2$ & \ding{51}\\
$\k$-Minkowski relative-locality $\N^2$ \cite{ArTr} (o.m.)   & 4   					& 3     					 & $+\infty$  				& 2		  & \ding{51}\\
Padmanabhan's non-local model \cite{ArCa1}	      				   & 4   					& 4     					 & $+\infty$  				& 2 & \ding{51}\\\hline\hline
\end{tabular}
\end{adjustwidth}
\caption{\label{tab1} Spacetime dimensions in various theories of quantum gravity with $D=4$ and for string theory in any target-spacetime dimension. References where the dimensions were calculated are indicated. In the case of non-commutative $\k$-Minkowski spacetime, different Laplacians $\N^2$ can be chosen, while ``o.m.'' and ``c.i.m.'' stand for, respectively, ordinary and cyclic-invariant spacetime measure. UV values for GFT/spin foams/LQG correspond to the smallest scales where spacetime can be approximated by a continuum. Theories with a near-IR parameter $\Gamma_{\rm meso}\gtrsim 1$ (see the text) are ticked.}
\end{table}
The calculation of some of these dimensions is derived from, but not in, the cited references and requires the following explanation.

The spectral and Hausdorff dimensions of GFT/spin foams/LQG were calculated numerically in \cite{COT3} for simplicial-complex quantum states of geometry, at all scales. However, the parameter $\G$ will be later employed only at scales where the continuum limit holds. Thus, the UV dimensions listed in the table for GFT/spin foams/LQG correspond to the smallest scales at which discreteness effects are negligible. Concentrating on the smallest scales of the continuum, the results of \cite{COT3} lead to the range $-3\leq\Gamma_{\rm UV}<0$, which essentially agrees with and encompasses the findings of several phenomenological continuum approximations of LQG. Perturbations of the so-called weave states, effective dispersion relations of the form \Eq{dr1} were found with $3/2\leq\b\leq3$, depending on the truncation scale \cite{GaPu,AMTU}, while the recovery of logarithmic corrections (calculated in the full theory \cite{KaMa}) in the entropy-area law of black holes leads to an effective dispersion relation with $\b=3$ \cite{ACAP}. In all these cases, the Hausdorff dimension of spacetime does not run and we can infer the value of $\Gamma=\dh/2-\b=2-\b$, comprised between $-1$ (UV limit) and $1/2$ (mesoscopic scales). Moreover, in \cite{MiTr} the UV value $\ds^{\rm UV}=1$ was reported for a continuum approximation, corresponding in our notation to $\Gamma_{\rm UV}=4/2-4/1=-3$. Therefore, the UV values $\Gamma_{\rm UV}=-1$ and $\Gamma_{\rm UV}=-3$ obtained in these continuum approaches lie within the interval $-3\leq\Gamma_{\rm UV}<0$ we obtain from \cite{COT3}.

In the case of string theory, $\dh=\dh^k=D$ in the low-energy limit. Here $D$ is the topological dimension of any string spacetime where the modified dispersion relation \Eq{dr1} holds up to some microscopic scale. This geometry could be the full target spacetime (e.g., $D=10$) or the bulk spacetime in small-scale compactifications \cite{Andriot:2017oaz} with compactification radius $O(\lp)$ (we do not consider compactifications with large extra dimensions or braneworld models which may give rise to a different effect; see also \Eq{Rc} and the discussion thereabout). The calculation of the spectral dimension may take different roads that, somewhat surprisingly, give the same result. The dimensional anomaly in a non-critical string setting induces a modification of dispersion relations of low-energy particles such that $\b=3/2$ at mesoscopic scales and $\b=2$ in the UV \cite{ACEMN}, so that $\Gamma_{\rm UV}=D/2-D/2=0$. On the other hand, the $D=10$ target spacetime generated by the critical string is endowed with a spectral dimension that cannot be smaller than the intrinsic dimension of the probe one uses to scan the spacetime geometry. This probe is the worldsheet, so that $\ds^{\rm UV}=2$ at scales below the string tension \cite{CaMo1} and, again, $\Gamma_{\rm UV}=0$.

In $\k$-Minkowski spacetime, metric observables and the notion of Hausdorff dimension are not straightforward to define and they admit inequivalent realizations. Here we either assume   an ordinary spacetime measure (o.m.) $\rmd^4x$ or we use the cyclic-invariant spacetime measures (c.i.m.) $\rmd^4x\,|{\bf x}|^{-3}$ \cite{AAAD} or $\rmd^4x\,|x_1x_2x_3|^{-1}$ \cite{ACOS}, whose Hausdorff dimension was analyzed in \cite{ACOS}. The momentum-space Hausdorff dimension was calculated in eq.\ \Eq{dhkmink}. The expression of the spectral dimension in the case of the relative-locality Laplacian can be written in integral form but we will omit it, since the case of the bicross-product Laplacian is very similar and can be treated analytically. The spectral dimension for that theory is \cite{ArTr}
 $\ds(\ell)=(8\k^2\ell^2+6)/(2\k^2\ell^2+1)$, so that, using \Eq{dhkmink},
\be\label{ga1}
\Gamma(\ell)=2-\frac{\dh^k(\ell)}{\ds(\ell)}=2-\frac{3}{2}\frac{2(\k\ell)^2+1}{[4(\k\ell)^2+3]\{1-(\k\ell)^2+2(\k\ell)^4[\sqrt{1+(\k\ell)^{-2}}-1]\}}\,,
\ee
and $\G\simeq\G_{\rm UV}=3/2$ at short scales ($\k\ell\ll 1$). At large (near-IR) scales $\k\ell\gg 1$,
\be\label{Gkappa}
\Gamma_\textrm{meso}\simeq 1+\frac{5}{12(\k\ell)^2}=1+\frac{5}{96\pi}\frac{\lp^2}{\ell^2}\,.
\ee

In Padmanabhan's scenario, the density of microstates giving rise to the Bekenstein--Hawking entropy-area law is such that the dispersion relation of fields living in the vicinity of a black hole receives a non-polynomial modification and fields near the event horizon are effectively non-local \cite{Pad98,Pad99}. In turn, this dispersion relation generates a peculiar profile of the spectral dimension such that $\ds$ diverges at the Planck scale. In $D=4$ topological dimensions, the Hausdorff dimension in position and momentum space coincides with the topological dimension, $\dh=\dh^k=4$. The spectral dimension was calculated in \cite{ArCa1} and runs from infinity at an absolute lower length scale $\ell=\sqrt{4\pi}\lp$ to 4, so that
\be\label{ga2}
\Gamma(\ell)= 2-\frac{4}{\ds(\ell)}\,,\qquad \ds(\ell)=1+\frac{3\ell^2}{4\pi\lp^2}\left\{\psi\left(\frac{3\ell^2}{8\pi\lp^2}\right)-\psi\left[\frac{3}{2}\left(\frac{\ell^2}{4\pi\lp^2}-1\right)\right]\right\},
\ee
where $\psi(x)=\p_x\Gamma(x)/\Gamma(x)$ is the digamma function and $\ell> \sqrt{4\pi}\lp$. At small scales $\ell\ll\lp$, $\G\simeq\G_{\rm UV}=2$, while in the near IR ($\ell\gg\lp$) $\ds^\textrm{meso}\simeq 4+10\pi{\lp^2}/{\ell^2}$ and
\be\label{Gpad}
\Gamma_\textrm{meso}\simeq 1+\frac{5\pi}{2}\frac{\lp^2}{\ell^2}\,,
\ee
the same form as \Eq{Gkappa}.


\section{Luminosity distance and GW amplitude in quantum gravity}\label{ludis}

In quantum gravity, \Eq{dLGR} can be modified in two ways. The first is a change in the way we measure time and spatial intervals, due to the deformation of the fabric of spacetime and of the clocks and rods living in it. This leads to a modification of geodesics, of the line element, or of other basic quantities of general relativity. Depending on the theory, diffeomorphism invariance and Lorentz invariance may be broken at certain scales and recovered in the IR. The second type of modification is of the dynamics, so that the Hubble parameter $H(z)$ differs from the one in GR. Overall,
\be\label{dL}
d_L^\textsc{em}=(1+z)\int_0^z\frac{\rmd z\,v(z)}{H(z)}\,,
\ee
where $v(z)=v[t(z)]$ is a non-trivial measure weight that encodes quantum-gravity effects on the cosmic clock. Obviously, one could include $v$ in an effective $\tilde H:=H/v$, but this is not convenient for the discussion below. 

First, let us concentrate on the choice of $v$. There are several quantum-gravity or beyond-Einstein-gravity scenarios with a non-trivial $v(z)$.


\subsection{Varying Hausdorff dimension only}\label{ludis1}

A general theorem \cite{first,revmu} states that, for \emph{any} theory on a continuous spacetime (or admitting a continuous spacetime in a certain asymptotic limit) where the Hausdorff dimension is scale dependent, its profile $\dh(t,\ell)$ is given uniquely by a parametrization which, surprisingly, also coincides with the dimension profile of deterministic and random multi-fractals. (This and the following statements hold also for the spectral dimension $\ds$ \cite{first,revmu}.) Each theory will fix the parameters to specific values, thus giving rise to a characteristic dimensional profile which, in some cases, can be distinguished from the profiles in other theories. The general parametric profile is an infinite series in terms of the observation scale but, without any loss of generality, only the first two terms are important for phenomenology. In particular, the time-dependent part of the measure is
\be\label{vt}
v(t)=1+\left|\frac{t}{t_*}\right|^{\a-1}F_\om(t)\,,
\ee
where $t_*$ is the characteristic time scale at which Lorentz invariance is broken and the multi-fractal nature of spacetime geometry becomes apparent (typically, $t_*\gtrsim \tp$) and $\a>0$ is a parameter related to the Hausdorff dimension below $t_*$ by
\be\label{dhst}
\a=\frac{\dh^*}{D}
\ee
if all directions have the same anomalous scaling; otherwise, $\a$ is the Hausdorff dimension of time only. In the IR, $\ds\simeq D$, while in the UV, $\dh\simeq \a\,D$. Depending on whether $\a$ is smaller or greater than 1, the dimension in the UV can be smaller or greater than $D=4$. Typically, $0<\a<1$. The harmonic function $F_\om$, where $\om$ is a parameter not to be confused with the GW frequency, is related to a discrete scaling symmetry in the deep UV and to a fuzzy stochastic structure at ultra-small scales; it can be set to 1 for the time being.

It is worth noting that the profile \Eq{vt} is \emph{model-independent} inasmuch as the only assumption made is that the Hausdorff dimension of time changes with the scale. The fundamentals and the dynamics of the theory, whatever they are, determine the values of the parameters $\a$, $t_*$, $\om$, and so on. Of course, there are cases where $t_*=\infty$ and there is no time-like dimensional flow.

To parametrize $v(t)$ in terms of the redshift, we assume a profile for the scale factor. We will consider two cases, a power law and de Sitter-type time evolution:
\be
\textrm{case 1:}\qquad a=\left(\frac{t}{t_0}\right)^p\,,\qquad \textrm{case 2:}\qquad a=\exp[H_0 (t-t_0)]\,.
\ee
Then, for $F_\om=1$
\ba
\textrm{case 1:}\qquad v(z)&=& 1+ \left(\frac{t_*}{t_0}\right)^{1-\a} (1+z)^{\frac{1-\a}{p}}\,,\label{pro1}\\
\textrm{case 2:}\qquad v(z)&=& 1+ (H_0 t_*)^{1-\a}|H_0t_0-\ln(1+z)|^{\a-1}\,.\label{pro2}
\ea
Note that the prefactor $(t_*/t_0)^{1-\a}$ can be extremely small if $\a\lesssim 1/2$. Plugging these expressions into \Eq{dL}, one gets a modification of the luminosity distance due to a running Hausdorff dimension in the time direction, and that can be used in observations of high-$z$ objects, such as in LISA. This modification, which could be studied analytically depending on the profile $H(z)$, is valid for any theory where $\dh^{\rm time}\neq 1$ independently of the dynamics, except when the factor $v$ cancels exactly with the modification in $H$ (we will not consider this possibility \cite{frc17}). To get some insight, we can take a regime where $H\simeq H_0\simeq 1/t_0$ is approximately constant. The $H\simeq H_0$ approximation is incompatible with the assumption of a power-law scale factor, unless one considers low-redshift sources. In this case, 
\be
\textrm{case 1:}\qquad d_L^\textsc{em} \stackrel{z\ll 1}{=} \frac{z}{H_{\rm eff}}+\frac{1+c_{1}(H_0 t_*)^{1-\a}}{H_0}z^2+O(z^3)\,,\label{c1}
\ee
where $c_1 = (2p+1-\a)/(2p)$ and 
\be\label{Heff}
H_{\rm eff}:=\frac{H_0}{1+(H_0 t_*)^{1-\a}}\,.
\ee

On the other hand, for the de Sitter case we find for all $z$
\be
\textrm{case 2:}\qquad d_L^\textsc{em} \simeq \frac{1+z}{H_0}\left(z+(H_0 t_*)^{1-\a}\{\Gamma[\a,1-\ln(1+z)]-\Gamma(\a,1)\}\rme\right),\label{dex2}
\ee
where $\Gamma(\a,x)$ is the upper incomplete gamma function and $\rme=\exp 1$ is Nepero number. To see whether we can use the parametrization \Eq{eq:param}, we expand \Eq{dex2} at small $z$. Up to a numerical constant, the result is the same as the asymptotic expression for case 1, eq.~\Eq{c1} with $c_1$ replaced by an $\a$-dependent coefficient $c_2$ which is always positive and $O(1)$ if $0<\a<1$.

The luminosity distance as determined by a nearby GW source can be approximated by \Eq{dz1}, with $H_0$ replaced by $H_{\rm eff}$. Identifying \Eq{c1} with the luminosity distance $d_L^\textsc{gw}$ determined by a source of gravitational waves, and noting that in GR for constant $H\simeq H_0$ the luminosity distance is \Eq{dz1}, one immediately gets
\be\label{cas1}
\Xi(z\ll 1)\simeq 1+(H_0 t_*)^{1-\a}+(c_{1,2}-1)(H_0 t_*)^{1-\a}\,z\,.
\ee
This function does not match the parametrization \Eq{eq:param} due to the extra constant factor $(H_0 t_*)^{1-\a}$. However, the latter is very small if $0<\a<1$. The present age of the universe is about $H_0^{-1}\simeq t_0\approx 14\times 10^9\,{\rm yr}\approx 10^{17}\,{\rm s}$, while the scale $t_*$ is much smaller. The Planck scale $\tp\approx 10^{-43}\,{\rm s}$ is a reasonable lower bound, while as an upper bound we can take a time scale below electroweak processes, as suggested by phenomenology \cite{revmu}, but above the grand-unification scale: $10^{-43}\,{\rm s}<t_*<10^{-30}\,{\rm s}$. Within this range,
\be\label{corre}
(H_0 t_*)^{1-\a}\sim (10^{-60}-10^{-47})^{1-\a}.
\ee
Neglecting this term, we can roughly identify $(c_{1,2}-1)(H_0 t_*)^{1-\a}$ with $(\Xi_0-1)n$, but the former is much smaller that the latter. In fact, a LISA estimate for the term $(\Xi_0-1)n$ is $O(10^{-2})$ \cite{BDFM}, with an error $O(10^{-3})$ (see also \cite{Belgacem:2019pkk} for a more recent estimate).  On the other hand, the term $(c_{1,2}-1)(H_0 t_*)^{1-\a}$ is roughly given by \Eq{corre}, further suppressed by a small number $c_{1,2}-1$. Even taking $c_{1,2}-1 = O(1)$, to reach the $O(10^{-3})$ level of the error on $(\Xi_0-1)n$, $\a>0.94-0.95$, a value too close to unity to be realistic in these scenarios. Thus, the effect will be too small to constrain the UV dimension of time (or of spacetime, if \Eq{dhst} holds and all directions have the same anomalous scaling).

Things change completely if we consider scenarios where $\a>1$. Theoretically, they would correspond to spacetimes where the Hausdorff dimension increases in the UV. We do not have concrete quantum-gravity theories with $\dh^{\rm UV}>D$ and we should regard this case as phenomenological (i.e., not top-down). For these values of $\a$, the coefficient $c_1$ may become negative, but still $|c_1|=O(1)=c_2$, and the factor \Eq{corre} can be very large if $\a\gtrsim 2$, $H_{\rm eff}\simeq H_0 (H_0 t_*)^{\a-1}\gg H_0$, which gives a dominant constant correction in \Eq{cas1} well within the LISA range. Apart from not being based upon any known quantum-gravity model, this result has the disadvantage of relying on a fixed (but general enough) profile of the scale factor $a(t)$. This means that the actual number $c_1$ or $c_2$ will depend on such a choice. Still, for a reasonable choice of $p$ an observational constraint on $\a$ could be found. Given that we have no support from known QG proposals, and that later on we will replace \Eq{cas1} with the more general and robust expression \Eq{dla2}, we will not insist on this possibility here.

Usually, the modification \Eq{dL} of standard cosmology does not entail any screening mechanism and local measurements of the effective Newton constant do not differ from cosmological observations. All propagating degrees of freedom, photons and gravitational waves alike, are affected in the same way by a change in spacetime measure, so that $\Xi=1$. However, this does not imply that $M_\textsc{flrw}=M_{\rm eff}=M_\textsc{gw}$ because the luminosity distance $d_L^\textsc{em}=d_L^\textsc{gw}$ is not proportional to the local effective Planck mass $M_{\rm eff}$, the reason being that effective Planck masses typically scale as a certain power of $v$, while in \Eq{dL} $v$ is integrated. Examples are the multi-fractional theories \cite{revmu}:
\begin{itemize}
\item In the case with so-called weighted derivatives, $M_\textsc{gw}^{D-2}=M_\Pl^{D-2} v$, $G_\textsc{flrw}=G/v$ \cite{frc17} (i.e., $M_\textsc{flrw}^{D-2}=M_\textsc{gw}^{D-2}=M_\Pl^{D-2} v$) and $M_{\rm eff}=M_\Pl$, so that
\be
\De M_\textsc{flrw}=0\,,\qquad \De M_{\rm eff}(z)=\frac{1}{v(z)}-1\,.
\ee
\item In the case with so-called $q$-derivatives, $M_\textsc{gw}^{D-2}=M_\Pl^{D-2} v/v^2=M_\Pl^{D-2}/v$, $M_\textsc{flrw}=M_\Pl$ and $M_{\rm eff}\sim M_\Pl v$. In the first expression, the factor $v$ comes from the action measure, just like for the theory with weighted derivatives, while the factor $1/v^2$ comes from the kinetic term. The second expression stems from the fact that the first Friedmann equation is $H^2/v^2=\rho/(3M_\Pl^2)$ \cite{revmu}, but there is another factor $1/v^2$ hidden inside $\rho$. To get the third expression, one notices that the Newton potential scales as $\Phi\sim -G/q^{D-2}(r)=1/(M_{\rm eff}r)^{D-2}$, where $q(r)\simeq \int^r\rmd r\, v(r)\sim r\,v(r)$. Thus,
\be
\De M_\textsc{flrw}(z)=v(z)-1\,,\qquad \De M_{\rm eff}(z)=v^2(z)-1\,,
\ee
valid for any $z$. Here, $\Xi=1$.
\end{itemize}
Since $v\to 1$ at late times (small redshift), locally all mass parameters $\De M$ vanish, while at large redshift they deviate from zero and, depending on the theory, can take different signs.


\subsection{Varying Hausdorff and spectral dimension}\label{ludis2}

How do photons and GWs propagate in spacetimes with dimensional flow? To answer this question, we first consider the relation between the electromagnetic luminosity distance and proper distance. Recall that we defined $d_L^\textsc{em}$ in \Eq{ludi} from the measurement of the flux and the power of an optical source. However, when written in terms of $r$ \Eq{ludi} is a Newtonian law that would break down at scales where dimension is not four, since areas, distances and potentials acquire an anomalous scaling. Recall that, in a multi-scale and/or fuzzy spacetime such as those arising in quantum gravities \cite{NgDa,Ame94,ACCR}, the measurement of a generic distance $r$ in the absence of curvature in a non-relativistic regime (Newtonian approximation) deviates from the one in ordinary space by a power-law correction, so that\footnote{The expression written in \cite{ACCR} has an extra denominator $\a$ in front of the correction, and is therefore ill-defined when $\a=0$. To allow also for this value, we redefined the coefficient, which is not strictly fixed by any law.} \cite{NgDa,Ame94,ACCR,first}
\be\label{L}
r\to\tilde r = r\left[1+\e\left(\frac{r}{l_*}\right)^{\a-1}\right],
\ee
exactly of the polynomial type typical of multi-scale systems. The parameter $\e$ accounts for two types of corrections. If $\e=1$, the effect is \emph{deterministic} and interpreted as a systematic correction to distance measurements, due to performing them in a spacetime with an anomalous geometry. According to the flow-equation theorem \cite{first,revmu}, when measuring $r$ with physical rods one can identify $\a$ with the UV Hausdorff dimension of spacetime divided by $D$ \cite{ACCR}. Generic quantum-gravity arguments select the values
\be\label{aval}
\a=\frac13\,,\,\frac12\,,
\ee
as especially appealing \cite{NgDa,Ame94,ACCR}. Alternatively, the correction in \Eq{L} may also be regarded as a \emph{stochastic} noise generated by the continuous movement of quanta of spacetime due to Heisenberg's indeterminacy principle. Then, in average $\langle\e\rangle=0$ and $\langle\e^2\rangle=1$, and the correction is an intrinsic measurement uncertainty, in which case spacetime is said to be \emph{fuzzy}.

The fundamental scale $l_*$ is the ideal divide between the IR and the UV. Spacetime geometry can have more characteristic scales, but $l_*$ is the largest one around which QG effects dominate. While theoretical arguments imply that $l_*$ is of order of the Planck scale $\lp$, a more phenomenological attitude is to let observations speak for themselves and constrain $l_*$. Upper bounds range from the electroweak to the grand unification scale (see \cite{revmu} and references therein). In this paper, the length $l_*$ will be assumed to be microscopic, $l_*=O(\lp)$. This has an important consequence for the optical luminosity distance. At scales where the cosmological expansion is negligible, the quantity $d_L^\textsc{em}$ in \Eq{ludi} is replaced by $\tilde r$, keeping in mind that the distance one measures is $r$, not $\tilde r$. At cosmological scales, redshift factors appear as in standard GR and one replaces $r$ in \Eq{ludi2} with $\tilde r$, so that the final expression for the electromagnetic luminosity distance is
\be\label{ludi3}
\boxd{d_L^\textsc{em} =\frac{a_0^2}{a}\tilde r=\frac{a_0^2}{a}\,r\left[1+\e\left(\frac{r}{l_*}\right)^{\a-1}\right].}
\ee
However, at large scales well above Planck, already above the electroweak interaction length, the QG effect in \Eq{ludi3} is totally negligible and \Eq{ludi2} is adequate for measurements above the laboratory scale, such as those considered in this paper. The case of a stochastic spacetime in the implementation of \cite{ACCR} is even simpler, since for long distances $\langle\e\rangle=0$ and \Eq{ludi2} holds to all purposes.\footnote{A modified dispersion relation with random fluctuations can elicit particle creation if one treats the changes in the dispersion at a quantum-field-theory effective level. For instance, heavily constrained vacuum Cherenkov radiation occurs in Lorentz-violating extensions of the Standard Model \cite{LePo,Alt07,AnTa}. However, these modifications of the Standard Model do not find an immediate, clear implementation in the quantum gravities discussed here, especially those that, as far as GWs are concerned, could give the most interesting signal (GFT/spin foams/LQG; see below).}

Another, equivalent way to obtain \Eq{ludi3} is through the tools of fractal geometry used in one specific model with an explicit running of the Hausdorff dimension, the so-called theory with $q$-derivatives \cite{first,revmu}. There, the flat FLRW line element is of the form $\rmd s^2\simeq-v^2(t)\,\rmd t^2+a^2(t)\,v^2(r)\,\rmd r^2=-v^2(t)\,\rmd t^2+a^2(t)\,\rmd\tilde r^2$, where $\tilde r(r)=\int^r\rmd r'\,v(r')$ and the approximation holds when changing from Cartesian to radial coordinates when the spacetime measure $v(t,{\bf x})$ is factorized in the $D$ coordinates. Then, for light rays $\rmd s^2=0$ and the multi-scale comoving distance \Eq{L} is given by $\tilde r(t)=\int_{t_0}^t \rmd t'\,v(t')/a(t')$. Reinstating the redshift factors in the definition of the optical luminosity distance, we obtain equivalence between \Eq{dL} and \Eq{ludi3}. In general QG, this line element and the calculation derived from it are an effective approximation describing Hausdorff dimensional flow, which agrees with the more traditional heuristic reasoning leading to \Eq{L} \cite{NgDa,Ame94,ACCR}.

We now turn to gravitational waves. The action \Eq{Svqg} for a scalar is replaced by an effective\footnote{Not in the sense of the one-loop quantum effective action.} second-order action for the tensor perturbation which is \Eq{hboxh}, with a time-dependent mass $M_\textsc{gw}(t)$, in the case of most phenomenological modified-gravity models, while in quantum gravity it can be of the form
\be\label{hboxh2}
\de S=\frac12\int\rmd^Dx\,v(x)\,\sqrt{-g}\,M^{2\Gamma}_\textsc{gw}\,h_{ij}\cK(\B_\textsc{flrw})h^{ij}\,,
\ee
where the mass $M_\textsc{gw}$ is constant\footnote{An effective non-constant mass $\tilde M^{D-2\b}_\textsc{gw}:=v\,M^{2\Gamma}_\textsc{gw}$ can be simply defined by absorbing the measure weight.} and it is elevated to the power $2\Gamma=\dh-2\b$ to make the action dimensionless. The equivalent of the scalar mode is the combination $\bar h:=M_\textsc{gw}^{\Gamma} h$. It is possible to write a modification of the Mukhanov--Sasaki equation \Eq{wgr} for the graviton polarization mode, with $w$ given by a rescaling of $\bar h$ just like in \Eq{wgr}, but with $[\bar h]=(D-2)/2$ replaced by \Eq{vpuv}, so that
\be\label{Newton}
h\propto \frac{1}{(M_\textsc{gw} r)^\G}
\ee
in the local wave zone. Note that this is also proportional to the Newtonian potential, which is the asymptotic solution of the Green equation $\cK(\N^2)\,h_{00} (r)=\de(r)$ on a static background. An example of calculation confirming \Eq{Newton} in a specific but fairly general model can be found in the Appendix.

Since the $r\to ar$ rule holds on the FLRW background independently of the dynamics, we do not need to know the details of the modified perturbation equation and the luminosity-distance dependence can be read off directly from the scaling law
\be\label{hh0}
w\stackrel{{\rm UV/meso}}{\simeq} (a M_\textsc{gw})^{\g}h\,,\qquad w\stackrel{{\rm IR}}{\simeq} (a M_\textsc{gw})^{\frac{D-2}{2}}h\,,
\ee
where
\be
\g=\G_*=\G_{\rm UV/meso}
\ee
is the scaling parameter \Eq{vpuv} in either a UV or a mesoscopic regime. The gravitational-wave amplitude scales as (superscript \textsc{em} omitted)
\be\label{hh}
h \propto \frac{1}{d_L^{\Gamma}}\qquad\Longrightarrow\qquad h \stackrel{{\rm UV/meso}}{\sim} \frac{1}{d_L^{\g}}\,,\qquad h \stackrel{{\rm IR}}{\sim} \frac{1}{d_L^{\frac{D-2}{2}}}\,.
\ee
Notice that this formula is derived under the assumption that $\G\approx{\rm const}$, which is valid only in any plateaux in dimensional flow, where dimensions run very slowly. This assumption breaks down away from these plateaux and we cannot trust \Eq{hboxh2} pointwise.

In order to combine the UV and the IR regimes of \Eq{hh} into a single, compact formula, one can consider two routes.

\smallskip
\noindent
{\bf First route:} As a first possibility, 
we  promote $\G$ to a scale-dependent parameter and write the transcendental relation
\be\label{hqm}
h\stackrel{?}{\propto} \frac{1}{d_L^{\Gamma(d_L)}}\,.
\ee
For a given theory, one should pick the expression for $\G(\ell)$ such as \Eq{ga1}, \Eq{ga2} or the profiles in Fig.\ \ref{fig3}, and replace $\ell$ with $d_L$. However, this procedure is heuristic and not well justified. As we just pointed out, the relation \Eq{hh} stems from a fairly general scaling argument applied to the Mukhanov--Sasaki variable, valid in a continuous spacetime and in a range of scales where the spacetime dimensions are approximately constant. In other words, $\G$ in \Eq{hh} is a constant that takes different values at different scales. In contrast, it is difficult to guess what type of Mukhanov--Sasaki equation the strain \Eq{hqm} would have to satisfy. To put it in a schematic way, the dimensionless action $S=({\rm meas})\times({\rm kin})$ is given by the product of the measure times the kinetic term. The measure is always a generalized polynomial $({\rm meas})\sim A_1 r^{a_1} + A_2 r^{a_2}+\dots$, where we work in the local wave zone. Also the kinetic term is a generalized polynomial in the local wave zone approximation \Eq{lwfa}, $({\rm kin})\sim B_1 r^{b_1} + B_2 r^{b_2}$, where the first term is the leading one in the $\om r\gg 1$ expansion. Keeping only the first two terms, we can arrange the length scale or scales inside $A_{1,2}$ and $B_{1,2}$ such that the four quantities $A_1B_1 r^{a_1+b_1}$, $A_1B_2 r^{a_1+b_2}$, $A_2B_1 r^{a_2+b_1}$ and $A_2B_2 r^{a_2+b_2}$, as well as their combination $S$, are dimensionless. However, in the case of \Eq{hqm} we would have to replace the kinetic term with something of the form $({\rm kin}') = C(r)\, r^{c(r)}$, where the functions $C(r)$ and $c(r)$ are constructed such that $C\to B_1$ and $c\to b_1$ in a certain (UV or IR) regime and $C\to B_2$ and $c\to b_2$ in another (IR or UV) regime. But then, one can never compensate the scaling of $({\rm meas})$, which is a rigid binomial, at \emph{all} scales. To summarize, the scaling of $h$ is determined by the varying elements of the action, i.e., the measure and the kinetic term. However, in the local wave zone approximation these elements have a well-defined scaling (or ``critical exponent'') only in plateaux, not in transition regimes (see, for instance, \Eq{dh}, \Eq{dr1} and \Eq{vt}). In particular, it would not make sense to replace $\G$ in \Eq{hboxh2} with a scale-dependent version $\G(\ell)$ because this effective exponent would \emph{not} compensate the rigid monomial, binomial or polynomial multi-scaling of the measure and the kinetic term pointwise.

There is another complication on top of that. The exact expression of the Green function of $h$ is strongly model-dependent and, more often than not, the spatial momentum integral, which must be performed before going to the local wave zone $\om r\gg 1$, cannot be calculated exactly, as we have seen in the Appendix where a simple modified dispersion relation already required to limit the Green function to a plateau of dimensional flow ($\cK\sim k^{2\b}$). Everything immediately extends to curved spacetime and to expressions of the luminosity distance. Therefore, the great majority of QG systems are such that \Eq{hqm} cannot be found exactly and, anyway, is not correct in the local wave zone approximation.

\smallskip
\noindent
{\bf Second route:}  A second option for obtaining
 a luminosity-distance formula valid both in the UV and in the IR is based on the observation that the asymptotic behaviour of Green functions with multiple anomalous scalings is quite general. Multi-scale systems as diverse as those in the fields of multi-fractal geometry, chaos theory, transport theory, financial mathematics, biological ecosystems, cognitive psychology and artificial intelligence display correlation functions with anomalous scalings described by the same mathematics. These systems are characterized by two or more critical exponents $\de_1$, $\de_2$, and so on (corresponding to dimensions, in QG and fractal geometry) combined together as a generalized polynomial $\ell^{\de_1}+ A\ell^{\de_2}+\dots$, where $A$ and each subsequent coefficient contains a scale. As far as we know, this is the standard result and there is no need nor evidence for replacing a finite set of critical exponents $\de_1$, $\de_2$, \dots, with a one-parameter exponent $\de(\ell)$ and the polynomial with a single power law $\ell^{\de(\ell)}$. 

Indeed, in the case of the luminosity distance, the length formula \Eq{L} is precisely of the polynomial form expected in multi-scale systems and it can provide a guidance to rewrite $d_L^{\G}$ in \Eq{hh} as the sum of an IR and a UV contribution. Reinstating the superscript in $d_L$,
\be\label{dla}
\boxd{h \propto \frac{1}{d_L^\textsc{gw}}\,,\qquad d_L^\textsc{gw}=d_L^\textsc{em}\left[1+\ve\left(\frac{d_L^\textsc{em}}{\ell_*}\right)^{\g-1}\right],\qquad \g\neq 0\,,}
\ee
\be\label{dlag0}
\boxd{h \propto \frac{1}{\ell_{*}} \ln\left(1+\frac{\ell_*}{d_L^\textsc{em}}\right),\qquad \g= 0\,,}
\ee
where the parameters $\ve=O(1)$, $\g$ and $\ell_*\geq l_*$ will be discussed shortly. First, we comment on the range of validity of \Eq{dla} when $\g$ takes values far away (say, 50\% or more) from 1. Assume, then, that $|\g-1|\geq 0.5$. Equation \Eq{dla} captures the scaling of the GW amplitude on two different regimes, one where $d_L^\textsc{gw}\simeq d_L^\textsc{em}$ (IR/GR regime, negligible correction) and one where $d_L^\textsc{gw}\simeq \ve\ell_* (d_L^\textsc{em}/\ell_*)^\g$ (UV/QG regime, dominant correction). Depending on the magnitude of $\g$, one regime corresponds to the scale of the observer, while the other to cosmological scales arbitrarily far away from us. If $\g<1$, then the GR regime is realized for optical sources with $d_L^\textsc{em}\gg\ell_*$, while if $\g>1$ it is realized when $d_L^\textsc{em}\ll\ell_*$. Whether the GR regime corresponds to cosmological or local (i.e., solar system, laboratory or atomic) scales depends on how dimensional flow affects the cosmological observable \Eq{dla}. Ultimately, this question reduces to determining whether $\g=\G_{\rm UV}$ or $\g=\G_{\rm meso}$. The magnitude of the QG correction in \Eq{dla} can change considerably depending on the regime and on the geometry.
\begin{itemize}
\item $\bm{\g=\G_{\rm UV}}$. A binomial such as \Eq{dla} is valid at all scales only if $\ell_*$ is the only intrinsic scale in spacetime geometry, in which case $\ell_*$ is expected to be very small, certainly smaller than the electroweak scale and possibly close, or equal, to the Planck length $\lp$. Therefore, for $\ell_*=l_*=O(\lp)$, $\g=\G_{\rm UV}$ is the critical exponent in the UV and cosmologically distant sources ($d_L^\textsc{em}\gg\ell_*$) fall into the IR regime of dimensional flow (GR limit) if $\g<1$ and into the UV regime (QG limit) if $\g>1$. Thus, interesting deviations from GR are expected only when $\g>1$. Note that we cannot conclude, from this reasoning, that at sub-cosmological scales (solar system, laboratory, and so on) one reaches the UV regime if $\g<1$ and the IR regime if $\g>1$, because \Eq{dla} is a cosmological formula and $d_L=0$ corresponds to zero redshift or local scales, not sub-Planckian scales. In particular, a theory with $\g<1$ does not necessarily predict strong QG effects at solar-system or laboratory scales. The case $\g=0$ is special because it corresponds to a logarithmic correlation function. Equation \Eq{dlag0} reproduces this behavior in the UV, $h\sim -\ln(d_L^\textsc{em}/\ell_*)$ when $\ell_*/d_L^\textsc{em}\gg1$, while in the IR $h\sim 1/d_L^\textsc{em}$ when $\ell_*/d_L^\textsc{em}\ll1$.
\item $\bm{\g=\G_{\rm meso}}$. Anomalous lengths such as \Eq{L} can be derived on general grounds by invoking a universal IR feature of dimensional flow in quantum gravity (namely, that the spacetime dimension flows to $D$ in the IR smoothly and slowly, through a plateau) \cite{first}. This means that \Eq{L} or its many-scale generalizations arise as a sort of perturbative expansion of the length around the IR, where perturbation terms are corrections with anomalous scalings. This IR expansion has as many terms as the number of fundamental scales $\ell_*>\ell_1>\ell_2>\dots$ of the geometry. Since this number is finite, then the IR expansion is also finite, hence it is exact and captures the whole dimensional flow. In the case of \Eq{dla}, we have used a scaling argument which compares the behaviour of a field at two different scales. If there is only one scale $\ell_*$ in the hierarchy, then one of these two scales is in the deep UV and $\g=\G_{\rm UV}$, but if $\ell_*$ is only the largest scale in an extended hierarchy, then the smallest scale can be a mesoscopic reference instead of an ultra-short one. Even in the absence of more than one fundamental scale, one can model the luminosity distance in the mesoscopic regime of dimensional flow by an expansion around the average distance $\ell_*$ of the sources observed in an experiment, assuming that they fall into the near-IR regime of dimensional flow (shaded region in Fig.\ \ref{fig3}). Then \Eq{dla} would be an effective formula in an expansion near the IR, where the correction to the luminosity distance is the leading term. In this case, $\g=\G_{\rm meso}$, we expect that $\ell_*\gg\lp$ and the discussion in the previous item must be revised. In fact, for a cosmologically large $\ell_*$ the regime $d_L^\textsc{em}\ll\ell_*$ may still correspond to very distant objects, falling into the IR/GR regime if $\g>1$ and into the meso/QG regime if $\g<1$.
\end{itemize}
So far, we have assumed that $\g$ is away from 1. However, there is one model in Tab.\ \ref{tab1} where $\g=\G_{\rm UV}=1$ exactly ($\k$-Minkowski spacetime with ordinary measure and bicovariant Laplacian) and another where $\g=\G_{\rm meso}\gtrsim 1$ in the near-IR regime (GFT/spin foams/LQG; see Fig.\ \ref{fig3}). The GR limit is not recovered in \Eq{dla} in these cases, unless $\ve$ is made $\g$-dependent and vanishing when $\g=1$. This choice is possible, since the scaling argument leading to \Eq{dla} does not determine uniquely the prefactor in front of the correction. If $\ve$ is a fixed number (deterministic spacetimes), the correction is systematic. Its sign is model dependent, while if $\g\neq 1$ its absolute value is 1 because there already is a free parameter $\ell_*$ governing the size of the quantum correction. In other words, since also $\ell_*$ is a free parameter, one can set the coefficient to be $\ve=O(1)$ without loss of generality. In this case, $\ell_*=l_*$. However, while in \Eq{L} the choice $\a=0$ is excluded by construction, in \Eq{dla} it may be possible that $\g\approx 1$, in which case the choice $\ve=\pm 1$ does not lead to the correct GR limit. This phenomenological reason implies that $\ve$ must have a $\g$ dependence. A simple choice such that $\ve(\g=1)=0$ could be $\ve=\pm(\g-1)^n$ for some positive $n$, but we can further restrict it by requiring that $\ve(\g\neq 1)=O(1)$ for the values of $\g=\Gamma_{\rm UV}$ listed in Tab.\ \ref{tab1}, hence $n=O(1)$. Moreover, we also want to recover the pure power law \Eq{hh} on any plateau of dimensional flow with $\g=\G=O(1)$, which selects $n=1$ uniquely. Therefore, for deterministic models we can set $\ve=\pm(\g-1)$.

In fuzzy spacetimes with intrinsic measurement uncertainties, $\ve$ is a random variable averaging to zero and with variance $\pm 1$. Then, the power-law correction in \Eq{dla} is an ``error'' of quantum-gravity origin, determining an upper threshold on a measurement precision. Therefore, for fuzzy models we allow for a sign ambiguity which, combined with the previous choice of $\ve$, gives an \emph{Ansatz} encompassing both the deterministic and the fuzzy scenarios:
\be\label{veg}
\ve=\pm|\g-1|\,.
\ee
This choice gives viable phenomenology for any value of $\g$, both when $|\g-1|=O(1)$ and when $|\g-1|\ll 1$. Moreover, \Eq{veg} allows us to refine our very broad definition of mesoscopic scales. Since $\g=1$ corresponds to the ideal IR limit, in the mesoscopic interpretation of \Eq{dla} we can define $\ell_*$ as the scale at which we get a transition from $\Gamma=1$ to values slightly away from 1 (near-IR limit). Exotic effects may show up around $\ell_*$, which acts as a sort of observational screen for any other scale below it \cite{revmu,first}. In this multi-scale case, $\ell_*$ is just an effective mesoscopic scale modeling dimensional flow towards its IR end and, depending on the model, it can stay close to the Planck length or be far away from it. On the other hand, the speed of dimensional flow at the observation scales is determined by the value of $\Gamma_{\rm meso}$. Since the flow is always sufficiently slow at these cosmological scales in the mesoscopic interpretation of \Eq{dla}, we conclude that $\Gamma_{\rm meso}$ should be very close to 1. If, on top of that, $\Gamma_{\rm meso}\gtrsim 1$, one could get a moderately large QG correction in \Eq{dla} for $d_L\gg \ell_*$.\footnote{This effect could be achieved also with a very mild dimensional flow such that $\G_{\rm UV}\gtrsim 1$, implying $\ds^{\rm UV}\gtrsim \ds^{\rm IR}=D$, so that the spectral dimension is only slightly larger than the topological dimension at short scales. From Tab.\ \ref{tab1}, we do not have concrete examples at hand, since the quantum gravities the authors are aware of with just one fundamental scale, and such that $\ds^{\rm UV}> \ds^{\rm IR}$, have a gap $\ds^{\rm UV}- \ds^{\rm IR}\geq 1$.}

To summarize, we have four cases:
\begin{enumerate}
\item[1.] $\bm{\g=\G_{\rm UV}<1}$. Here $d_L^\textsc{em}\gg\ell_*=O(\lp)$ and the correction in \Eq{dla} is negligible at all cosmological scales. The theories in Tab.\ \ref{tab1} falling into this case are GFT/spin foams/LQG, causal dynamical triangulations, $\k$-Minkowski spacetime with cyclic-invariant measure and bicovariant, bicross-product, or relative-locality Laplacian, Stelle gravity, the low-energy limit of string theory, asymptotic safety, and Ho\v{r}ava--Lifshitz gravity. This category is among the most interesting from the point of view of high-energy physics. Let $\dh^{\rm UV}=\tilde{d}_{\rm W}^{\rm UV}$. For $\dh^k=\dh$, one has $\g=\G_{\rm UV}=0$ when $\dh^{\rm UV}=0$ (a UV value corresponding to a degenerate spacetime) or $\ds^{\rm UV}=2$ (the ``magic number'' sometimes advertized \cite{tH93,Car09,fra1}, or even over-emphasized, as a sign of renormalizability in quantum gravity). Then, the GW amplitude is scale-invariant in the UV. The luminosity distance receives an $O(\ell_*)$ correction (or becomes fuzzy at scales $<\ell_*$, if we interpret the correction term as a stochastic noise with amplitude $\ell_*$). Since the GW signal to which present and near-future interferometers are sensitive was generated by macroscopic objects and propagated on cosmic distances, this UV regime is most likely unobservable in these experiments. An alternative would be to find phenomenological models with a spectral dimension $\ds\simeq 2$ in the IR rather than in the UV but, to the best of our knowledge, there is no such model available in quantum gravity.
\item[2.] $\bm{\g=\G_{\rm meso}\lesssim 1}$. The correction in \Eq{dla} is negligible for cosmological sources with $d_L^\textsc{em}\gg\ell_*\gg\lp$ and may become moderately important for closer sources at $\lp\ll d_L^\textsc{em}\lesssim\ell_*$. The theories in Tab.\ \ref{tab1} falling into this case are causal dynamical triangulations, $\k$-Minkowski spacetime with cyclic-invariant measure and bicovariant, bicross-product or relative-locality Laplacian, Stelle gravity, the low-energy limit of string theory, asymptotic safety, and Ho\v{r}ava--Lifshitz gravity. Since none of these theories display more than one characteristic scale, $\ell_*\gg\lp$ is purely phenomenological and corresponds to the cosmological scale below which dimensional-flow effects become manifest. Note that this is not the exact analogue of case 1 because it does not include GFT/spin foams/LQG, for which $\G_{\rm UV}<1\lesssim\G_{\rm meso}$. See Fig.\ \ref{fig3}, where mesoscopic scales correspond to the local maximum above 1 before the fall off towards the IR.
\item[3.] $\bm{\g=\G_{\rm UV}>1}$. Here $d_L^\textsc{em}\gg\ell_*=O(\lp)$ and the correction in \Eq{dla} can be important at all cosmological scales because, contrary to case 1, here $\g-1>0$. The theories in Tab.\ \ref{tab1} falling into this case are $\k$-Minkowski spacetime with ordinary measure and bicross-product or relative-locality Laplacian, and Padmanabhan's non-local model near a black hole.
\item[4.] $\bm{\g=\G_{\rm meso}\gtrsim 1}$. The correction in \Eq{dla} is negligible for cosmological sources with $\lp\ll d_L^\textsc{em}\ll\ell_*$ and may become moderately important for farther sources at $d_L^\textsc{em}\gg\ell_*>\lp$. The theories in Tab.\ \ref{tab1} falling into this case are GFT/spin foams/LQG for $\ell_*>\lp$, $\k$-Minkowski spacetime with ordinary measure and bicross-product or relative-locality Laplacian for $\ell_*\gg\lp$ (the theory does not predict an intermediate scale, hence $\ell_*$ is cosmological), and Padmanabhan's non-local model for $\ell_*\gg\lp$. This is not equivalent to case 3 because $\G_{\rm UV}<1\lesssim\G_{\rm meso}$ in GFT/spin foams/LQG, where $\G_{\rm UV}<1$ increases up to a peak at $\G_{\rm meso}>1$ just before reaching the asymptote $\G_{\rm IR}=1$. In Fig.\ \ref{fig3}, the scale $\ell_*$ at which $\G_{\rm UV}<1$ is about forty times larger than the Planck scale but its actual position is state-dependent. In fact, GFT/spin foams/LQG are the only theories in Tab.\ \ref{tab1} where $\ell_*$ is an effective scale encoded in a quantum state of geometry. This scale is larger than $\lp$ but does not have to be as large as in case 2. To check this in GFT/spin foams/LQG, one should construct quantum states of geometry such that $\Gamma_{\rm meso}\gtrsim 1$ along the lines of the numerical analysis of \cite{COT3}, and verify that the parameter space of these states is physically sensible. We will postpone this to the future and conclude, for the time being, that $\ell_*>\lp$ for GFT/spin foams/LQG, while $\ell_*\gg\lp$ for $\k$-Minkowski spacetime and Padmanabhan's non-local model. These examples are ticked in Tab.\ \ref{tab1}.
\end{enumerate}

Equation \Eq{dla} is a profile which we \emph{inferred} from the relation \Eq{hh} \emph{derived} with a model-independent scaling argument, but we did not derive \Eq{dla} directly from the Mukhanov--Sasaki equation of the gravitational perturbation because, in general, it is not possible to do it exactly. However, it is not sensitive to the effective background mass in the Mukhanov--Sasaki equation, which is negligible with respect to the kinetic term within the Hubble horizon. The background $a(t)$ may or may not differ substantially from that of GR, but here we do not use this information. The main effect encoded in \Eq{dla} stems from the scaling of the wave-amplitude solution. We expect this equation to hold for any theory with a modified dispersion relation (varying $\ds$) and/or a running $\dh$, in the local wave zone approximation.

Note that the deterministic version of \Eq{dla} (i.e., when $\ve$ is a fixed number) is very similar to the one in cosmological braneworld models with extra dimensions and a cross-over scale $R_{\rm c}$ at which the four-dimensional dynamics receives corrections from the bulk \cite{DeMe}:
\be\label{Rc}
h \propto \frac{1}{d_L^\textsc{gw}}\,,\qquad d_L^\textsc{gw}=d_L^\textsc{em}\left[1+\left(\frac{d_L^\textsc{em}}{R_{\rm c}}\right)^{n_{\rm c}}\right]^\frac{D-4}{2n_{\rm c}},
\ee
where $n_{\rm c}>0$ governs the transition steepness. From \Eq{Rc}, one can put constraints on $R_{\rm c}$ and on the topological dimension of the bulk upon detection of standard sirens \cite{Nis17,PFHS,Abb18}. Note that a major difference with respect to \Eq{dla} is that \Eq{Rc} is completely \emph{ad hoc}, in the sense that it is inspired by \Eq{hcos} but it does not follow the polynomial
behaviour universally found in mathematical as well as natural multi-scale systems. On the other hand, \Eq{dla} follows this generalized polynomial behaviour, present also in QG formul\ae\ such as \Eq{L}.

Although there is much freedom about what models underlie \Eq{dla}, we can concentrate on the data-driven task of placing model-independent constraints on the parameters $\g$ and $\ell_*$. The estimated error on the luminosity distance in the combination of LIGO, Virgo and KAGRA is about $\De d_L/d_L\sim 0.2$ \cite{DHHJ,NHHDS}. In the case of the Einstein Telescope, the accuracy on the $d_L$ measurement is $\De d_L/d_L\sim 0.001-0.01$ for $z\ll 1$ and $\De d_L/d_L\sim 0.1$ for $z>2$ \cite{CaNi}. LISA will have similar figures \cite{Tam16,Belgacem:2019pkk}. For DECIGO, these estimates are about one order of magnitude smaller \cite{CaNi}. This means that the effect would be within the detectability window if
\be\label{bou31}
|\g-1|\left(\frac{d_L^\textsc{em}}{\ell_*}\right)^{\g-1}\gtrsim 10^{-3}-10^{-1}\,.
\ee
If $\g>1$, this condition could be met even when $d_L\gg \ell_*=\lp$. Below we will confirm the bound \Eq{bou31} when $\g$ is close to 1.

Before concluding the section, let us discuss the parameter \Eq{dez3}. From \Eq{dla},
\be\label{dla20}
\Xi(z)=1+\ve\left[\frac{d_L^\textsc{em}(z)}{\ell_*}\right]^{\g-1}.
\ee
In order to know $d_L^\textsc{em}(z)$, one should pick specific theories of quantum gravity and check from the model-dependent dynamics \Eq{dL} how photons propagate on a cosmological background. In general, some theories may modify the electromagnetic sector, while others do not (in which case $d_L^\textsc{em}$ is standard). In this paper, we do not pretend to be exhaustive, since we are mainly interested in some model-independent features and in setting some general lines for future investigation on the topic of gravitational waves in quantum gravity. For this reason, and for simplicity, we will not attempt a complete treatment of the problem and will only consider cases where $d_L^\textsc{em}$ is unaffected by QG. At small redshift,
\be\label{dla2}
\Xi(z\ll 1)\simeq 1+{\ve}\left(\frac{z}{H_0\ell_*}\right)^{\g-1}\,.
\ee
The fact that the spectral dimension runs in \Eq{dla2} but not in \Eq{cas1} explains the discrepancies in their form. On one hand, when $\ds=D$ and $\dh=\dh^k$, $\g=\dh^*(D-2)/(2D)=\a(D-2)/2$, so that in $D=4$, and since $\ell_*=t_*$ in Planck units, \Eq{dla2} becomes $\Xi\simeq 1+\ve (H_0t_*)^{1-\a}z^{\a-1}$, which still disagrees with \Eq{cas1} by a constant correction and the leading-order redshift dependence. However, a running spectral dimension is associated with a modified dynamics (hence a modified profile $H(z)$), contrary to expression \Eq{cas1} where $\ds=4$ and the Hubble parameter was left untouched. Given these differences, the more general expression \Eq{dla2} supersedes the one found in section \ref{ludis1}. We must stress, however, that we cannot assume that $H$ is constant for certain LISA standard sirens, such as super-massive black hole merging  at high redshifts (up to $z\sim5-10$). In these cases, the simple formula \Eq{dla2} does not apply and one might have to investigate the problem numerically from \Eq{dla20}.

The parametrization \Eq{eq:param} does not hold except when $\g=2$ ($2/\dh=1/2-1/\ds$), one of the special cases giving rise to a large signal. To allow for non-integer powers of $z$ instead of $1+z$, we suggest the alternative small-redshift parametrization
\be\label{param2}
\Xi(z\ll 1)= 1+(\Xi_0-1)\left(\frac{1-a}{a}\right)^m=1+(\Xi_0-1) z^m\,,
\ee
where $m>0$ and $\Xi_0$ is a real parameter. In contrast, the parametrization \Eq{eq:param} is always of the form $\Xi(z\ll 1)=1+O(z)$ at low redshift. A possible extension of \Eq{param2} to all redshifts is
\be\label{param3}
\Xi(z)= \Xi_0+(1-\Xi_0)\,\rme^{-z^m}\,,
\ee
which has the same number of free parameters and the same large-$z$ asymptotic limit as \Eq{eq:param}.\footnote{One may even consider a three-parameter generalization of \Eq{param3} $\Xi(z)= \Xi_0+(\Xi_1-\Xi_0)\,\rme^{-z^m}$ to get a non-unit small-$z$ asymptote. In general, however, a two-parameter formula will be more suitable to use with data, and one of these three parameters will have to be fixed by the theory.}

The parameter $\Xi$ depends on the photon dynamics and we do not have a model-independent expression for it. For the same reason, we cannot tell \emph{a priori} whether $m=\g-1$ in \Eq{param2} and \Eq{param3} or if the large-redshift limit of \Eq{param3} efficiently encodes quantum-gravity models. These questions may deserve further study in the future.


\subsection{Dimension in scalar-tensor and modified-gravity models}\label{dhost}

The scaling law \Eq{hh} came out from pure dimensional arguments, but we can see how it can be generated directly from the action \Eq{hboxh2} in the special case where all QG effects are confined to the background and the dispersion relation is not modified. Since this assumption is restrictive in general, it is meant to be valid only in a near-IR regime. Defining $w=\tilde a h$ with $\tilde a:=a^\G$ as in \Eq{hh0}, for a constant $\G$ one has $w''-(\tilde a''/\tilde a)w=\tilde a[h'' + 2(\tilde a/\tilde a) h']=\tilde a(h'' + 2\G\cH h')$, which indicates that the propagation equation resulting from \Eq{hboxh2} and giving rise to the scaling \Eq{hh} is
\be\label{mseff}
h''+2\Gamma\cH h'+O(h)=0\,,
\ee
where in the last term we included also the spatial kinetic part. This implies that the relation between our parameter $\g$, Nishizawa's time variation $\nu$ of the effective Planck mass \cite{Nis17} (equivalent to the parameter $\a_M$ \Eq{aM}) and the parameter $\de$ of \cite{BDFM} is $\dh-\tilde{d}_{\rm W}=D-2+\nu=D-2+\a_M=(D-2)(1-\de)$ in $D$ topological dimensions, implying
\be\label{gnude}
{\nu=\a_M=-(D-2)\de=2\G-(D-2)=(\dh-D)+(2-\tilde{d}_{\rm W})\,.}
\ee
Thus, we can reinterpret the time variation of the effective Planck mass from an effective
point of view as a relation between the Hausdorff and spectral or walk dimension of spacetime. In other words, the parameters $\nu$ and $\de$ tells us how much the Hausdorff and walk dimension deviate from their ordinary classical values $\dh=D$ and $\tilde{d}_{\rm W}=\dw=2$. Closer to the deep UV, in some theories the effective Mukhanov--Sasaki equation may be higher-order in time derivatives or even non-local, in which cases \Eq{mseff} may break down. Therefore, we expect \Eq{gnude} to be accurate only at mesoscopic scales where these derivative corrections are suppressed.

The result \Eq{gnude} gives an alternative perspective  into the geometry of spacetime in non-QG models of dark energy such as scalar-tensor theories (see, e.g., \cite{Lan17,KaTs} for recent reviews). We now prove that \Eq{gnude} holds also in these cases.

The non-minimal coupling between gravity and other degrees of freedom modifies the propagation and dispersion relation of gravitational waves, as one can appreciate from \Eq{hboxh}. The effect is encoded in a time-varying effective Planck mass $M_\textsc{gw}(t)$, which stems from a class of actions with a non-minimal coupling of the form  $M_\textsc{gw}^{D-2}(\phi,X)\,R$, where $R$ is the Ricci scalar, $\phi$ is a scalar field and $X=\N_\mu\phi\N^\mu\phi$. For Jordan--Brans--Dicke theory \cite{BrDi,Dic62}, $M_\textsc{gw}^{D-2}=\phi$; for scalar-tensor models, $M_\textsc{gw}^{D-2}=F(\phi)$ is a non-linear function of the scalar; for general scalar-tensor models, $M_\textsc{gw}^{D-2}=2\cG(\phi,X)$ is a covariant function of the scalar and its derivatives; and so on \cite{BeSa,GLV}.

Regardless of its field dependence, $M_\textsc{gw}$ gives rise to an effective, dynamical change of the Hausdorff dimension of spacetime. In fact, we can regard the factor $M_\textsc{gw}^{D-2}(t)$ as a modification of the spacetime effective measure in \Eq{hboxh} as seen by the tensor perturbation $h_{ij}$, in the measurement units of the Jordan frame \cite{Dic62} where gravity is non-minimally coupled with the scalar field. In contrast, higher-order or non-local derivative operators acting on $h_{ij}$ would modify its dispersion relation and, consequently, the spectral dimension as probed by gravitons.\footnote{The spectral dimension is a particle-dependent concept and different particle probes may experience different spectral dimensions, even in standard quantum field theory in flat space \cite{CMNa}.} While in modified-gravity models the dimensional flow of $\dh$ is driven by the dynamics and is essentially an IR effect, in quantum gravity it originates in a deep modification of the spacetime texture in the UV.

This being the case, the position-space Hausdorff dimension can be calculated as a function of the FLRW background dynamics. The effective $D$-volume as seen by the tensor perturbation is $\cV\propto\int\rmd^Dx\,\sqrt{-g}\,M_\textsc{gw}^{D-2}$. Note that the engineering dimensionality of $\cV$ is $[\cV]=-2$, but this is not the Hausdorff dimension because the latter comes from the scaling of the coordinate dependence (non-constant part) of $\cV$. Then,
\be\label{for1}
\dh=D+\frac{\rmd\ln M_\textsc{gw}^{D-2}}{\rmd\ln\ell}\,.
\ee
The first term comes from the ordinary  volume density $\rmd^Dx\,\sqrt{-g}$, while the second is a correction that can be evaluated once the dynamics is solved. The generic scale $\ell$ can be replaced by a more suitable time scale, for instance the scale factor or the redshift. The scale factor is the most appropriate choice, since it is in direct relationship with the metric coordinates. Therefore, $\ell \propto a$ and \Eq{for1} becomes
\be\label{for2}
\boxd{\a_M=\nu=\dh-D\,.}
\ee
On the other hand, $\dh^k=D=\ds$ at all scales (the Fourier transform is the usual one with standard momentum measure) and the right-hand side of \Eq{gnude} is $(\dh-D)+(2-2)=\dh-D$, in agreement with \Eq{for2}. This result is possible because, in this class of models, the position-space and momentum-space Hausdorff dimensions are different, $\dh\neq\dh^k$.

The conclusion is that one can reconstruct the redshift dependence of the Hausdorff dimension of scalar-tensor theories and modified-gravity models after $\a_M$. In all these models, the parameter $\a_M=\nu$ receives a geometric interpretation as the difference between the effective Hausdorff dimension of spacetime (determined by the classical background dynamics, not by quantum effects) and the topological dimension.


\section{Observational constraints on quantum gravity}\label{nums}

In this section, we discuss the constraints that can be imposed on the QG theories that have been discussed above. In particular, we will first place bounds on possible deviations from GR in the measurements of the luminosity distance with GWs. We will consider two multi-messenger sources, the first binary neutron star (BNS) detected by the LIGO-Virgo collaboration, namely GW170817, and a simulated supermassive black hole (SMBH) binary detectable with LISA at high redshift. We will then discuss complementary constraints from testing the speed of propagation of GWs and consider possible bounds derived from solar-system observations.


\subsection{Constraints from gravitational-wave astronomy}\label{numan}

Multi-messenger GW events can be used to test deviations from GR in the propagation of GWs \cite{DeMe,PFHS,Abb18,BDFM}. In fact, if the amplitude $h$ does not scale as $1/d_L$ as GWs propagate through the universe, the value of the luminosity distance retrieved by a parameter estimation of the signal detected by an interferometer will differ from the value measured through EM observations of the same source (or its hosting galaxy). By comparing the two different measures of $d_L$, it is thus possible to constrain any deviation from GR, which predicts that they must have the same value exactly.
We can thus set constraints on the QG models discussed above, by considering data collected by such multi-messenger GW events, for which, in particular, one should be able to determine the luminosity distance from EM observations, beside measuring it with GWs.

We will consider two such events. First, we will use public data on GW170817 collected by the LIGO-Virgo collaboration, together with EM observations of its hosting galaxy NGC~4993, from which an EM measurement of the luminosity distance has been obtained. The GW measurement of GW170817 yields a luminosity distance of $40^{+8}_{-14}$ Mpc \cite{Ab17a}, while the EM value recovered by surface brightness fluctuation methods applied to NGC 4993 is $40.7\pm2.4$ Mpc \cite{Cantiello:2018ffy}. The comparison between these two distance measurements has already produced constraints on some beyond-GR gravitational theories \cite{PFHS,Abb18}. The second event that we will consider is a simulated SMBH binary merger detected by LISA. We choose the place of this event at $z=2$, where the majority of EM counterparts to LISA SMBHs are expected to be observed \cite{Tam16,Tamanini:2016uin}. Using standard $\Lambda$CDM cosmology with parameters adjusted to the Planck measurement \cite{P18I}, this redshift corresponds to a distance of $15.96$ Gpc. For this SMBH merging event, we will furthermore assume a relative 1$\sigma$ Gaussian error on the luminosity distance measured from GWs of $0.0614$, reflecting the expected systematic uncertainties due to weak lensing (LISA instrumental errors are usually dominated by weak lensing) \cite{Tam16}. We will also assume a 15\% absolute 1$\sigma$ error on the redshift measurement, which is expected to be obtained with photometric techniques at high redshift \cite{Tam16}.

Using the two multi-messenger GW events above, we will place constraints on the parameters $\g$ and $\ell_*$ in the expression \Eq{dla}. As upper or lower bounds on the parameters, we choose values corresponding to 90\% of the total allowed area of their posterior.

The Bayesian analysis we use is identical to the one of~\cite{Abb18} performed for the luminosity-distance relation~\eqref{Rc}. In our case, we investigate the luminosity distance relation~\eqref{dla} (and~\eqref{dlag0} for $\g=0$). For each case $\ve > 0$ or $\ve < 0$, we will fix one of the two parameters $(\g,\ell_{*})$, and put a constraint on the other by treating it as a derived parameter. For instance, when fixing $\ell_{*}$ and inferring a constraint on $\g$, we write the posterior distribution of $\g$ given the observed EM and GW data $(x_{\textsc{em}}, x_{\textsc{gw}})$ as
\be
	p(\g | x_{\textsc{em}}, x_{\textsc{gw}}, \ell_{*}) = \int dd_L^\textsc{em}\, dd_L^\textsc{gw}\, p(d_L^\textsc{em} | x_{\textsc{em}}) \,p(d_L^\textsc{gw} | x_{\textsc{gw}})\, \delta \left[ \gamma - \gamma(d_L^\textsc{em}, d_L^\textsc{gw}, \ell_{*})\right] \,.
\ee
The posterior for $\ell_{*}$ for a fixed $\g$ is obtained by exchanging the roles of $\g$ and $\ell_{*}$. Here the $p(d_{L} | x)$ factors are the EM and GW observed posteriors for the luminosity distance given the data (in the case of the GW observation, marginalized over all other parameters of the source). The second term in the Dirac delta is the expression for $\g$ as a function of ${d_L^\textsc{em}}$, $ {d_L^\textsc{gw}}$ and $\ell_{*}$ obtained by inverting~\eqref{dla}. This assumes that the map is univalued, an issue we will come back to below. In practice, we compute the integral above by using discrete posterior samples for the joint posterior $p(d_L^\textsc{em} | x_{\textsc{em}})\, p(d_L^\textsc{gw} | x_{\textsc{gw}})$, with domain such that $d_L^\textsc{gw} > d_L^\textsc{em}$. Note that treating the inferred parameter as a derived parameter means that we do not impose a prior on the inferred parameter itself, however priors on ${d_L^\textsc{em}}$ and ${d_L^\textsc{gw}}$ are included implicitly in their posterior distributions. We use the same approach for~\eqref{dlag0} where $\ell_{*}$ is the only parameter to constrain.

\begin{figure}
\centering
\includegraphics[width=15.8cm]{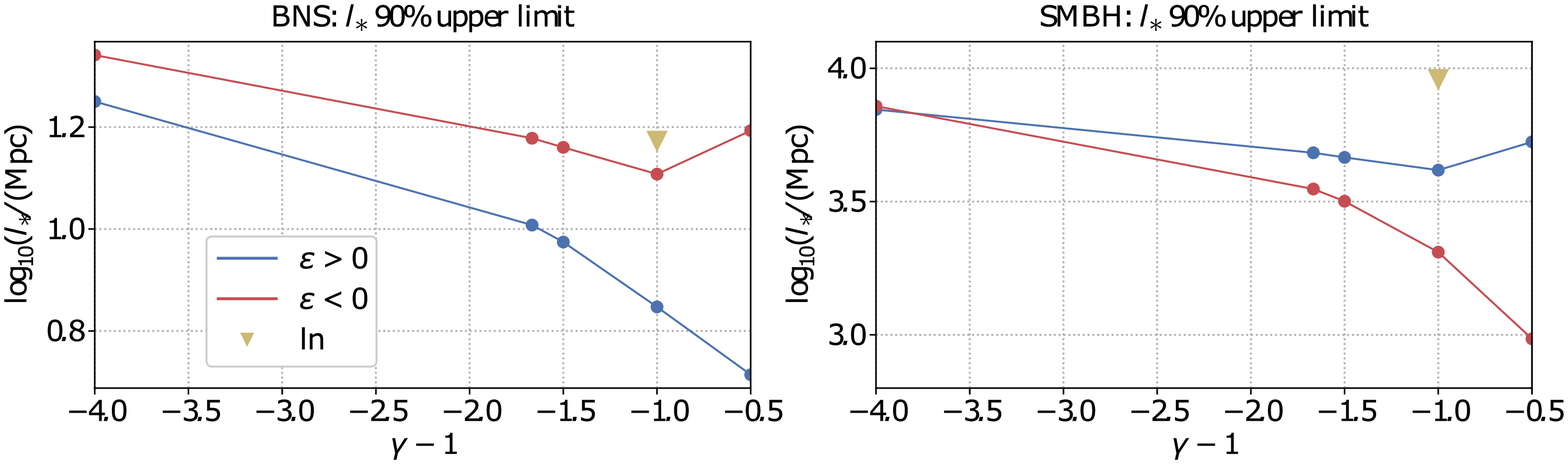}\\
\includegraphics[width=15.8cm]{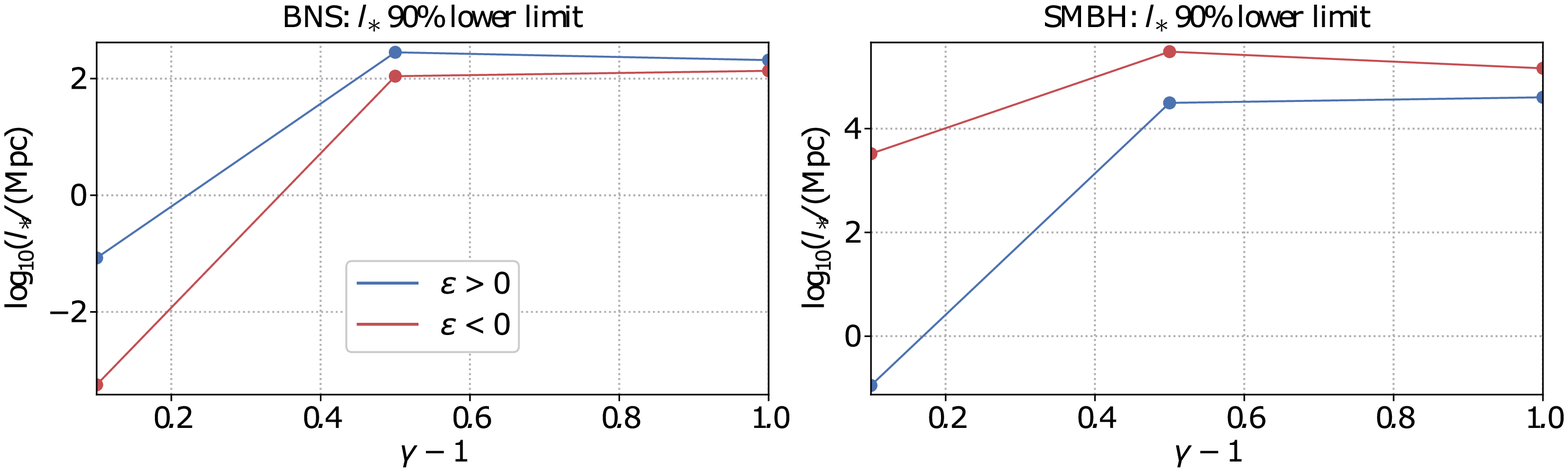}
\caption{\label{fig4} Upper bounds (top plots, for $\g < 1$) and lower bounds (bottom plots, for $\g > 1$) on $\ell_*$ at the 90\% confidence level, for the BNS and SMBH observations, as in Tab.~\ref{table2}. Blue and red show the two cases $\varepsilon > 0$ and $\varepsilon < 0$ in \eqref{dla}, while the yellow arrow shows the result for the logarithmic distance relation~\eqref{dlag0}.}
\end{figure}

First, we explore both the UV-limit and the near-IR regime of the theories listed in Tab.~\ref{tab1}. As we said, these theories were chosen mainly because their dimensional flow is well known. The cosmological phenomenology of some of them has not been explored in the literature, while for others (e.g., LQG; see section \ref{lqc}) there exist various inflationary models. Even in those cases where some phenomenology exists, the luminosity distance of gravitational waves has never been considered before, which means that the constraints we get are independent of any other. Fixing $\g$, we get an upper or a lower limit on the scale $\ell_*$, depending on the sign of $\g-1$. When \Eq{dla} is regarded as valid throughout the whole dimensional flow, we can pick the values in Tab.~\ref{tab1}, all falling in the interval $-4\leq\g-1\leq 1$. In this case, the scale $\ell_*$ may be identified with a fundamental (and possibly Planckian) scale of geometry. As one can see in Fig.\ \ref{fig4} and Tab.\ \ref{tab2}, the BNS gives tighter (respectively, weaker) constraints than the SMBH binary when $\g-1<0$ (respectively, $\g-1>0$). 
\begin{table}
\centering
\begin{tabular}{c|rr|rr}\hline
       & \multicolumn{2}{|c|}{BNS} & \multicolumn{2}{|c}{SMBH} \\ \hline
$\g$   & $\ve>0$ & $\ve<0$            & $\ve>0$    & $\ve<0$ \\\hline\hline
$-3$   & $18$ & $22$ & $7\times 10^3$ & $7\times 10^3$ \\
$-2/3$ & $10$ & $15$ & $5\times 10^3$ & $4\times 10^3$ \\
$-1/2$ & $9$  &	$14$ & $5\times 10^3$ &	$3\times 10^3$ \\
0 ($\ln$) & \multicolumn{2}{|c|}{$15$} & \multicolumn{2}{|c}{$9\times 10^3$} \\
$1/2$  & $5$  & $16$ & $5\times 10^3$ & $1\times 10^3$ \\ \hline
$1.1$  & $8\times 10^{-2}$ & $6\times 10^{-4}$ & $1\times 10^{-1}$ & $3\times 10^3$ \\ \hline
$3/2$	 & $3\times 10^2$    & $1\times 10^2$ & $3\times 10^4$ & $3\times 10^5$\\
2	     & $2\times 10^2$ & $1\times 10^2$ & $4\times 10^4$ & $1\times 10^5$\\ \hline\hline
\end{tabular} \\
\caption{\label{tab2} Upper ($\g<1$) and lower ($\g>1$) bounds on $\ell_*$ (in Mpc) at the 90\% level when $\g$ is fixed, for $\varepsilon > 0$ and $\varepsilon < 0$, using the luminosity distance relation~\eqref{dla}. We also inserted (with the label $\ln$) the result obtained with the logarithmic expression~\eqref{dlag0} replacing the case $\g=0$.}\label{table2}
\end{table}

Consider the top plots of Fig.\ \ref{fig4}. As expected ({\bf case 1. $\bm{\g=\G_{\rm UV}<1}$} in section \ref{ludis2}), when $\g-1<0$ all the upper bounds found for the UV values of Tab.~\ref{tab1} are cosmologically large and we conclude that this type of observations are unable to constrain the UV geometry of the following quantum gravities efficiently via the luminosity distance: GFT/spin foams/LQG, causal dynamical triangulations, $\k$-Minkowski spacetime with cyclic-invariant measure and bicovariant, bicross-product, or relative-locality Laplacian, Stelle gravity, the low-energy limit of string theory, asymptotic safety, and Ho\v{r}ava--Lifshitz gravity. $\k$-Minkowski spacetime with ordinary measure and bicovariant Laplacian is also unconstrained but for a different reason, since the luminosity distance is not modified there. On the other hand, in {\bf case 2. $\bm{\g=\G_{\rm meso}<1}$} we get a phenomenological upper bound for the scale $\ell_*$ below which dimensional flow is compatible with GW observations. The theories interested by this bound are all the above except GFT/spin foams/LQG.

When $\g-1>0$ (bottom plots of Fig.\ \ref{fig4}) we get a cosmologically large lower bound on $\ell_*$. If \Eq{dla} were valid at all scales, then we would be forced to rule out experimentally the UV limit of the three models of {\bf case 3. $\bm{\g=\G_{\rm UV}>1}$} ($\k$-Minkowski spacetime with ordinary measure and bicross-product or relative-locality Laplacian, and Padmanabhan's non-local effective model) as relevant at the cosmological scales probed by GWs. Of course, this does not exclude the possibility of residual corrections in the near-IR limit of these theories compatible with observations.

This would be {\bf case 4. $\bm{\g=\G_{\rm meso}>1}$}, which also includes GFT/spin foams/LQG (Fig.\ \ref{fig3}). Here we can get information about how fast the spacetime dimensions flow from the ideal IR (exactly four dimensions, infinitely large scales) to the scale of observation, which is by definition mesoscopic. Letting $\ell_*$ free and fixing $\g$, we should redo the above analysis with values of $\g-1$ slightly above 0. Here we only include the value $\g-1=0.1$ (Tab.\ \ref{tab2}), since the code fails to converge if $|\g-1|$ is too small. In this case, the lower limit on $\ell_*$ decreases and it reaches high-energy physics values for $\g-1\ll 0.1$. Also, in general, when $\g-1$ is away from 0 the sign of $\ve$ in \Eq{veg} is not particularly important, while for $\g-1\approx 0$ it may become crucial to determine the bound on $\ell_*$.

\begin{figure}
\centering
\includegraphics[width=15.8cm]{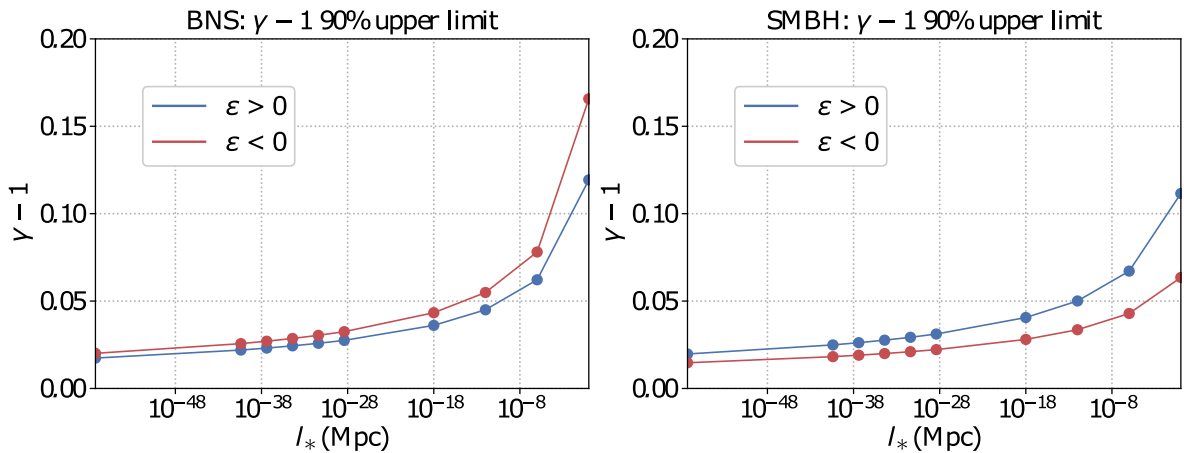}
\caption{\label{fig5} 
Upper bounds on $\g$ at the 90\% level, for $\ell_*$ fixed between $1\,{\rm Mpc}$ and the Planck scale $5\times 10^{-58}\,{\rm Mpc}$ and for both cases $\ve>0$ and $\ve < 0$.}
\end{figure}

To probe more in detail the $\g\approx 1$ regime, we fix $\ell_*$ and infer $\g$. The most interesting values of $\ell_*$ cover the energy scales probed in the LHC ($E_*\approx 10\,{\rm TeV}$, $\ell_*=10^{-20}\,{\rm m}=5 \times 10^{-43}\,{\rm Mpc}$) down to grand unification scales ($E_*\approx 10^{16}\,{\rm GeV}$, $\ell_*=10^{-33}\,{\rm m}=5 \times 10^{-55}\,{\rm Mpc}$) and the Planck scale ($E_*\approx 10^{19}\,{\rm GeV}$):
\be\nonumber
\lp=10^{-35}\,{\rm m}=5 \times 10^{-58}\,{\rm Mpc}\,.
\ee
The results for $\ell_*$ ranging from $1\,{\rm Mpc}$ down to the Planck scale are shown in Fig.\ \ref{fig5} and Tab.~\ref{tab3}. The relation between $d_L^\textsc{gw}$ and $d_L^\textsc{em}$ can be bivalued for $\g$, while the formalism we use treats the inferred parameter $\ell_*$ as a derived parameter, which requires to write a well-defined function $\g(d_L^\textsc{gw},d_L^\textsc{em})$. To alleviate this problem, we added the constraint $\g>1$ and limited ourselves to the range $\ell_*<d_L^\textsc{em}$, enough to hold for all samples.
\begin{table}
\centering
\begin{tabular}{rc|rr|rr}\hline
      & & \multicolumn{2}{|c|}{$\ve>0$} & \multicolumn{2}{|c}{$\ve<0$} \\ \hline
$\ell_*/{\rm Mpc}$ & $E_*$ & BNS   & SMBH           & BNS    & SMBH \\\hline\hline
$1$			 					 &       & 1.119 & 1.112 & 1.166 & 1.063 \\ 
$1 \times 10^{-6}$ & 			 & 1.062 & 1.067 & 1.078 & 1.043 \\
$1 \times 10^{-12}$ &			 & 1.045 & 1.050 & 1.055 & 1.034 \\
$1 \times 10^{-18}$ &			 & 1.036 & 1.041 & 1.043 & 1.028 \\
$4 \times 10^{-29}$ & 1 eV & 1.028 & 1.031 & 1.032 & 1.022 \\
$4 \times 10^{-32}$ & 1 keV& 1.026 & 1.029 & 1.030 & 1.021 \\
$4 \times 10^{-35}$ & 1 MeV& 1.024 & 1.028 & 1.029 & 1.020 \\
$4 \times 10^{-38}$ & 1 GeV& 1.023 & 1.026 & 1.027 & 1.019 \\
$4 \times 10^{-41}$ & 1 TeV& 1.022 & 1.025 & 1.026 & 1.018 \\
$5 \times 10^{-58}$ & $\ep$& 1.017 & 1.020 & 1.020 & 1.015 \\\hline\hline
\end{tabular}
\caption{\label{tab3} Upper bounds on $\g$ at the 90\% level when $\ell_*$ is fixed, for both the BNS and the SMBH observation and for the two cases $\ve >0$ and $\ve<0$.}
\end{table}
As one can see, the upper bound on $\g-1$ saturates to about a 2\%:
\be\label{gabo}
\Gamma_{\rm meso}-1<0.02\,.
\ee
This is compatible with the estimate \Eq{bou31}, since the left-hand side reduces to $\sim|\g-1|$. For $\dh\simeq\dh^k\approx 4$, the bound \Eq{gabo} translates into the bound \Eq{1} for the spectral dimension.

We can now discuss the models of quantum gravity which predict $\Gamma_{\rm meso}-1\gtrsim 0$ at mesoscopic scales. Replacing $\ell$ with $d_L$ in the asymptotic expressions \Eq{Gkappa} and \Eq{Gpad} for $\Gamma_{\rm meso}$ in $\k$-Minkowski and Padmanabhan's model and plugging in the range of values $d_L\sim 10-10^2\,{\rm Mpc}$ for LIGO sources, we get
\be\nonumber
\Gamma_{\rm meso}-1\sim 10^{-121}-10^{-116}\,,
\ee
way below the bound \Eq{gabo}. For the LISA source at $d_L\approx 16\,{\rm Gpc}$,
\be\nonumber
\Gamma_{\rm meso}-1\sim 10^{-125}-10^{-124}\,.
\ee
Therefore, the IR regime of these models leaves no imprint in GW propagation.

The case of GFT/spin foams/LQG is less obvious and more promising because the spacetime dimension is strongly state-dependent and, despite the fact that area and volume minimal sizes are of order of the Planck scale, the effective length scale encoded in quantum geometry states defining macroscopic near-classical geometries may be much larger than the Planck length $\lp$. It remains to be seen whether realistic quantum states could produce a dimensional flow such that $\Gamma_{\rm meso}-1=O(10^{-2})$.

All the above results assume that the luminosity distance of electromagnetic sources is not affected by quantum gravity. This is an \emph{ad hoc} assumption that should be checked by actually computing the propagation of light in these theories. Future research may have to focus on this important detail before drawing any definite conclusion on the observability of QG models.


\subsection{Bounds from propagation speed}\label{othe}

As we have seen in section \ref{dires}, dimensional flow is also influenced by modifications of the dispersion relation \Eq{dr1}, which can be written as
\ba
\cK(-k^2)&=&-\ell_*^{2-2\dh^k/\ds}k^2+k^{2\dh^k/\ds}\nonumber\\
&=&-\ell_*^{-n}k^2\left[1-(\ell_*k)^{n}\right]\,,\qquad n=\dh-2-2\G\,,\label{dr12}
\ea
where we used \Eq{dsgen}. The LIGO-Virgo black-hole merging events have been used to place constraints on this modified dispersion relation, using the observation that the group velocity of the gravitational wave must be very close to the speed of light \cite{YYP,EMNan,ArCa2}. However, the limits obtained in this way are weaker than the ones we have found here because the GW frequency is much lower than the Planck frequency. Hence, one gets very weak bounds on $\ell_*$ that leave the theory virtually unconstrained. In alternative, setting $\ell_*$ to be no bigger than the LHC scale, one has $n<0.76$ \cite{ArCa2}. In GFT/spin foams/LQG, the Hausdorff dimension hits the IR limit at shorter scales than the spectral dimension, so that we can set $\dh^{\rm meso}\approx 4$, in which case
\be
\G_{\rm meso}>0.62\,.
\ee
This constraint is complementary to \Eq{gabo} but, in practice, it does not give new information.


\subsection{Solar-system tests}\label{compa}

Gravitational-wave observations and measurements of the luminosity distance can offer new probes of dimensional changes in the gravity sector, independently of other measurements and possibly complementary to those. This is not too dissimilar in spirit from what happens in theories with large extra dimensions, where only gravity can leak (while the Standard Model fields are confined on a brane), as in the DGP model recently constrained by LIGO \cite{Abb18}. Also those models predict distinctive cosmological effects (for instance, in the non-linear growth of structure) that can constrain the number of dimensions, but GW physics offers independent and perhaps much cleaner observables to constrain the number of extra dimensions.

To compare our findings on the theories which drew more attention in our analysis (i.e., those with $\G_{\rm meso}\gtrsim 1$, ticked in Tab.\ \ref{tab1}) with observations at other scales, first we recapitulate what we know so far. The phenomenology of $\k$-Minkowski and Padmanabhan's black-hole model have not been explored in much detail. Non-commutative inflation (section \ref{ncc} and \cite{CQC}) shares the same general framework of non-commutative spacetimes as $\k$-Minkowski but it is not quite the same model. Here we gave a small contribution to this topic at the scales of non-primordial GWs, ruling out any role in GW propagation of the UV regime of these models while concluding that the IR regime leaves no trace. For GFT/spin foams/LQG there are many more results available, the greatest majority in the early universe (section \ref{lqc} and \cite{CQC}). However, the constraints we got from GWs are independent of any other, let them be from inflation or, less frequently, from dark energy.

The reader might find it hard to accept the chance that a theory of quantum gravity modifying the gravitational physics at short scales could leave a imprint at cosmological scales while being compatible with other observations at, say, solar-system or laboratory scales. Let us state right off the bat that this possibility is no big news: there exists ample literature on inflationary models such as loop quantum cosmology (section \ref{lqc}), non-commutative cosmology (section \ref{ncc}) and string cosmology (e.g., \cite{CQC} and references therein) where QG effects are stronger at the largest scales, while leaving other sectors of physics unmodified or affecting them in a viable way. The point is that, in order to leave a mark simultaneously at very small and very large scales, it is not necessary to invoke a miraculous recovery of GR at local scales where precision tests place the tightest constraints. Simply, the physical observables examined at each scale are different and constraints on one do not necessarily overlap with another. 

Having said that, we also admit that the case of inflation is somewhat more intuitive than that of GWs. The largest observable scales in the cosmic microwave background correspond to very short inflationary scales that reentered the causal horizon, while the smallest ones were sub-horizon and classical when perturbations were generated. In the case of GWs, we do not have this magnification of trans-Planckian scales to cosmic size.

 Let us consider laboratory or solar-system scales where redshift factors are negligible. We now argue that gravitational solar-system tests can place stronger constraints than GWs observations if the Newton potential is modified but the electromagnetic sector is not.

In perturbative QG, one can calculate quantum corrections to the Newton potential in a robust way. At one loop, the leading correction is quadratic:
\be\label{effdon}
\Phi\sim-\frac{1}{r}\left(1+\xi\frac{l_\Pl^2}{r^2}\right),
\ee
up to a classical relativistic correction. The sign and value of the coefficient $\xi$ depend on the framework and on the absence or presence of matter fields, getting slightly different results in effective field theory ($\xi=41/(10\pi)$ without matter fields) \cite{BBDH2,KhKi},\footnote{Other values and signs for $\xi$ were quoted in early papers (e.g., \cite{Do94a,Do94b,HaLi,DaMa}) but the result of \cite{BBDH2} is regarded as definitive.} from entropic holographic arguments \cite{MoRan} and from the effective average action in asymptotic safety ($\xi=43/(30\pi)$ without matter fields) \cite{SaCM}. The same effective potential \Eq{effdon} is obtained in the context of the AdS/CFT correspondence \cite{Gubser:1999vj}. All these approaches agree on the $O(r^{-2})$ behaviour of the correction, which can be understood with a dimensional analysis. Noting that loop diagrams involve an extra power of $\k^2\propto \lp^2$, in the presence of \emph{only one} fundamental scale $\lp$ the only dimensionless combination we can form is $\propto \lp^2/r^2$. This leading order modification in the low-curvature, low-energy IR limit of quantum gravity is strongly suppressed by the Planck scale and is therefore unobservable at solar-system scales.

In other QG approaches, corrections to the Newton potential take a different form. In quantum Regge gravity, at large distances the potential fitting the numerical simulations of lattice configurations of geometry is Yukawa-like \cite{HaWi}:
\be\label{effdon2}
\Phi\sim-\frac{\rme^{-mr}}{r},
\ee
where $m$ is an effective mass for the graviton. When the mass is very small, the Yukawa behavior is not distinguishable from a pure power law
\be\label{effdon3}
\Phi\sim-\frac{1}{r^\s},
\ee
where $\s>1$ is scale dependent. In asymptotic safety with resummed graviton propagator, the effective Newton potential is \cite{War02}
\be\label{resgravV}
\Phi\sim-\frac{1}{r}\left(1-\rme^{-c r}\right),
\ee
where $c$ is a constant. This expression is a one-loop result, but it is valid at all distances $r>1/c$, including near the Planck scale. The same formula is obtained exactly in three dimensions from the Fourier transform of a propagator $\sim -1/(k^2+ak^4)$, stemming from the renormalization-group-improved Laplacian found in the functional renormalization approach of the theory.

If we take \Eq{Newton} at face value as the effective Newton potential in theories with dimensional flow (an assumption we will reassess later), then we obtain a behaviour such as \Eq{effdon3} instead of the effective-field-theory one \Eq{effdon}. This discrepancy can be understood as follows. When considering dimensional-flow non-perturbative effects, extra scales may enter the picture and modify the low-curvature result \Eq{effdon}, since quantum corrections may now be combinations of different geometry scales, not all of them small \cite{BCT1}. Such is the case of GFT/spin foam/LQG quantum states of geometry giving rise to the profile in Figs.\ \ref{fig2} and \ref{fig3}. In this case, mesoscopic corrections where $\G>1$ are less suppressed than the perturbative $O(r^{-3})$ correction in \Eq{effdon} in the near-IR. 

To be more precise and write a scale-dependent expression for $\Phi$, we refocus the discussion to luminosity distances. Even in theories where the matter sector is untouched, dimensional flow may creep in when measuring distances, time intervals, particle lifetimes, and so on. Clocks and rulers are deformed in a spacetime with an anomalous Hausdorff dimension. The typical modification of these measurements, conducted in a flat regime, is of the type \Eq{L} \cite{NgDa,Ame94,ACCR}. In particular, the optical luminosity distance $d_L^\textsc{em}$ is given by \Eq{ludi3}, where redshift factors are equal to 1 at sub-cosmological scales. Applying \Eq{L} to modifications of particle processes, one can constrain $l_*$ \cite{ScM,ZS,MuS,CO,revmu,frc16}. For instance, the UV scale $l_*$ cannot be larger than $10^{-10}-10^{-20}\,{\rm m}$ (100 eV to 1 TeV), depending on the model, while for the values \Eq{aval} these bounds are stronger, reaching the grand-unification or the Planck scale. Regarding $\a$, related to the Hausdorff dimension in the UV, typical upper bounds are slightly weaker than the one we will present in section \ref{stn}, valid only for theories where $\dh$ runs: $\k$-Minkowski with cyclic-invariant measure and GFT/spin foams/LQG. In particular, the rough bound \Eq{dhuvbou} is above the value in $\k$-Minkowski and quite close to the one in GFT/spin foams/LQG, which means that it does not constrain these models significantly, nor does it overlap with the constraints on the more involved scaling parameter \Eq{vpuv}. Thus, experimental constraints of \Eq{ludi3} at sub-cosmological scales restrict the allowed values of $l_*$ and $\a$ (Hausdorff dimension), while experimental constraints on \Eq{dla} at cosmological scales restrict the values of $\ell_*$ and $\g$ (a combination of Hausdorff and spectral dimensions).

Concerning $d_L^\textsc{gw}$, plugging \Eq{ludi3} into \Eq{dla}, at sub-cosmological scales ($a=a_0=1$) we have
\be\label{dlast}
d_L^\textsc{gw}=l_*\left[\frac{r}{l_*}+\e\left(\frac{r}{l_*}\right)^\a+\ve\frac{\ell_*}{l_*}\left(\frac{r}{\ell_*}\right)^\g+O(r^{\a\g})\right].
\ee
In stochastic spacetimes, both corrections are zero in average at super-Planckian integrated distances, since $\langle\e\rangle=0=\langle\ve\rangle$. In deterministic spacetimes, let us set $\e=1$ and $\ve=\pm|\g-1|$. The first term always dominates over the second at super-electroweak scales, since $r/l_*\gg 1$ and $0<\a<1$. Therefore, from now on we can ignore the $\a$ term. Assuming that $\Phi\propto h_{00}$ obeys the same equation as $h_{ij}\sim 1/d_L^\textsc{gw}$ (see \Eq{Newton} and the related discussion), from \Eq{dlast} we can read the effective Newtonian potential in four topological dimensions and split it into the GR term and a correction:
\ba
\Phi&\propto& \frac{1}{r+\ve\ell_*(r/\ell_*)^\g}\simeq \frac{1}{r}\left[1-\ve\left(\frac{r}{\ell_*}\right)^{\g-1}\right]=:\Phi_{\rm GR}\left(1\mp\frac{\De\Phi}{\Phi}\right)\,,\label{New0}\\
\frac{\De\Phi}{\Phi}&=&|\g-1|\left(\frac{r}{\ell_*}\right)^{\g-1}\,.\label{Newpot}
\ea
When $\g<1$, the correction $\De\Phi/\Phi$ is completely negligible at distances $r\gg\ell_*$, regardless of whether $\g=\G_{\rm UV}$ or $\g=\G_{\rm meso}$. The case $\g>1$ is more delicate. In the UV regime of \Eq{dla}, where $\g=\G_{\rm UV}$ and $\ell_*=l_*$, $\De\Phi/\Phi\gg 1$ and it is therefore ruled out for being incompatible with observations. In the near-IR regime of \Eq{dla}, $|\g-1|=|\G_{\rm meso}-1|\ll 1$ but the ratio $\ell_*/l_*$ could be large in principle, if the scale $\ell_*$ at which $\G_{\rm meso}\gtrsim 1$ is much greater than the Planck length. In this case, the coefficient $|\g-1|\ell_*/l_*$ can be $O(1)$ and the third term in \Eq{dlast} slightly dominates over the first, unless $\G_{\rm meso}$ is severely fine tuned to be close to 1, as in the cases \Eq{Gkappa} and \Eq{Gpad} of $\k$-Minkowski and Padmanabhan's model. As we saw, these scenarios are unobservable in GW experiments. The only possibility standing after this screening is GFT/spin foams/LQG, where $\G_{\rm meso}\gtrsim 1$ can be close but not too close to 1 and, crucially, $\ell_*$ is larger but not much larger than $l_*$. GR tests within the solar system using the Cassini bound give a rough estimate of $\De\Phi/\Phi<(\De\Phi/\Phi)_{\rm obs}\approx 10^{-5}$ \cite{BIT,Wil14}. This bound assumes that light geodesics do not differ from GR at solar-system scales. Inverting \Eq{Newpot} with respect to $\g$, we find $\g-1=W[(\De\Phi/\Phi)_{\rm obs}\ln(r/\ell_*)]/\ln(r/\ell_*)$, where $W$ is the principal value of the Lambert function. The effect is maximized with $\ell_*=\lp$. Plugging this value and the distance of Cassini $r\sim 1\,{\rm AU}\approx 15\times 10^{10}\,{\rm m}$ at the time of measurement, we obtain
\be
\G_{\rm meso}-1<\left(\frac{\De\Phi}{\Phi}\right)_{\rm obs}\approx 10^{-5}\,.
\ee
This bound is stronger than \Eq{gabo} and, therefore, we conclude that solar-system tests of the Newtonian potential can be more effective than GWs to constrain quantum gravity. However, this does not mean that GW bounds are always weaker than those from GW astronomy. In some cases, gravitational waves bring unique information that cannot be replaced by other sources, as we will see in the next section. More importantly, our formalism captures some aspects of the effective dynamics but not all, and it is not obvious that one can always identify $h_{00}$ with the Newton potential. In QG, the derivation of the Newton potential can be highly non-trivial, to the point where we do not know its form in some theories. This is indeed the case for LQG/spin foams/GFT, where the classical Newton potential (actually, the graviton propagator) has been recovered \cite{BMRS} and there are arguments pointing towards the possibility that in the UV it is regularized \cite{CLS}, but the fact is that an expression for an effective $\Phi(r)$ has not been found yet. Another example where extrapolation of the Newton law from the effective luminosity-distance law is unclear is the renormalization-improved dynamics in the functional renormalization approach to asymptotic safety. In this approximation, coupling constants acquire a dependence on the renormalization-group scale, which in turn is identified with some specific length or momentum scale. The origin of the potential \Eq{effdon} can be understood intuitively (up to a negative sign) when considering that the Newton coupling $G(k)$ depends on the IR cut-off (renormalization-group scale) $k$ as $G(k)=G_0(1+G_0k^2/g_*)^{-1}$, where $G_0$ is the classical value and $g_*$ is its dimensionless value at the UV fixed point. Identifying the scale $k$ with the inverse length $r$, one gets $\Phi\sim-G(r)/r$ with $G(r)\simeq G_0[1-G_0/(g_* r^2)]$ at large scales. However, an identification of the renormalization-group scale with the square root of the Kretschmann invariant for the Schwarzschild metric yields $k\propto r^{-3/2}$ and a different (cubic) correction to the potential \cite{Held:2019xde}, while an identification with an energy-momentum scale leads to the logarithmic law \Eq{dlag0}. Thus, in this theory predictions are sensitive to the scale identification. An additional complication is that we used the renormalization-group-improved Laplace--Beltrami operator leading to the quartic ($\b=2$) dispersion relation \Eq{dr1} \cite{LaR5}. While this may be the full kinetic term for scalars, for gravitons there could be additional contributions discussed only recently that would make the Newton potential finite at $r\sim 0$ \cite{Bosma:2019aiu}, in line with the result \Eq{resgravV} in the resummed-propagator approach \cite{War02}. Other recent development might bypass the ambiguity of the renormalization-improved dynamics approximation \cite{Platania:2019kyx}.

Therefore, we cannot compare \Eq{New0} with any full, unambiguous calculation. Finally, we are also unaware of, but we cannot exclude \emph{a priori}, the existence of a screening mechanism in QG preserving the classical form of the Newton potential while allowing for cosmological modifications of the luminosity distance. To summarize, \Eq{New0} is valid provided
\begin{enumerate}
\item we can identify $h_{00}\propto\Phi$ with the Newton potential and
\item there is no screening mechanism at solar-system scales,
\end{enumerate}
but given that we are not in the position to guarantee the validity of these assumptions, nor to check \Eq{New0} against calculations in the full GFT/spin foams/LQG theory, the results of this section should be taken with a grain of salt. On top of that, we stress that in all cases where $\G<1$ and only one fundamental scale of geometry is present ($\lp$), the correction is negligible, compatibly with the effective field theory approach and with the analogous negative results we found for the GW luminosity distance.



\section{Specific models of quantum cosmology}\label{stqg}

In this section, we briefly review some scenarios of quantum gravity where the luminosity distance is modified. Loop quantum cosmology (LQC) and non-commutative inflation deserve a special mention, since both may be able to produce an observable signal in the LISA range of frequencies. On one hand, they can generate an inflationary blue-tilted tensor spectrum which is boosted at short scales, and that could be detectable as a primordial stochastic background. On the other hand, in these two models the luminosity distance of gravitational waves is modified with respect to the general-relativistic one and we can investigate whether the parametrization \Eq{eq:param} is sufficient to encode the effect. Unfortunately, but interestingly, the LQC model considered below is unobservable in the case of individual sources due to the LIGO constraint on the propagation speed. From now on, $D=4$.


\subsection{Loop quantum cosmology}\label{lqc}

Quantum-gravity theories have mainly implications in the UV regime, hence often recovering an early era of accelerated expansion while avoiding the introduction of an inflaton field with a fined-tuned potential (e.g., \cite{deCesare:2016axk,deCesare:2016rsf} in the context of group field cosmology). Here we addressed the question of whether such theories can leave an imprint in late-time GW astronomy through dimensional-flow effects. Among the examples we considered, we saw that the UV continuous regime of LQG, explored both with numerical simulations of simplicial-complex quantum states and effective dispersion relations, is outside our reach, while a near-IR regime may be more promising. However, there is a third way to explore this theory, namely, via one of its semi-classical cosmological incarnations.

LQC is a mini-superspace cosmological model of LQG that admits several versions, one of which is the so-called anomaly-cancellation (or effective-constraints or effective-dynamics) scenario (see \cite{BBCGK} for a conceptual review). In this model, there are two types of quantum corrections to the cosmological dynamics of linear inhomogeneous perturbations: inverse-volume and holonomy corrections. When considering only the first, the Mukhanov--Sasaki equation for the tensor modes $w_k:= a_{\rm inv} h_k$ is \cite{BH2,BoC,CH}
\be\label{mukkh}
w_k''+\left[(1+2\a_0\dpl)k^2-\frac{a_{\rm inv}''}{a_{\rm inv}}\right]w_k=0\,,\qquad a_{\rm inv} :=  a\left(1-\frac{\a_0}{2}\dpl\right),
\ee
where $\dpl= (a/a_*)^{-\s}\sim |\tau|^{\s}\sim k^{-\s}$, $a_*$ is related to a state-dependent length scale, $\a_0>0$ and $0<\s\leq 3$. If $\s<1$, the tensor spectrum may be blue tilted in a certain regime \cite{BCT2,Zhu15,Zhu16}, thus giving rise to GW phenomenology. In fact, a blue-tilted spectrum compatible with CMB observations may be greatly enhanced at LISA frequency scales. Here we are interested in the luminosity distance and related parameters. \emph{A priori}, it is not obvious that the quantum correction is negligible, since $\dpl$ is the ratio of a background energy (length) scale over a quantum-state-dependent energy (length) scale that can be much smaller (larger) than the Planck scale.

The friction term for $h = a_{\rm inv}^{-1} w$ is $2\p_\tau a_{\rm inv}/a_{\rm inv}=\cH(2+\a_0\s\dpl)$, implying that
\be\label{lqcd}
\dh-\tilde{d}_{\rm W}-2=\nu=\a_M=-2\de=\a_0\s\dpl>0\,.
\ee
This relation, previously unnoticed, links the dimension of spacetime calculated for realistic quantum states with the cosmology of these models. However, the left-most hand side should be evaluated carefully. In the deep UV, but also at mesoscopic scales, the Hausdorff dimension in LQG is $\dh^{\rm meso}\gtrsim\dh^{\rm UV}=2$, while $\dw=2$ at all scales \cite{COT3}. At the scales of the underlying simplicial complex where discreteness effects dominate, the spectral dimension is zero, but at mesoscopic scales where discreteness blends in a continuum, $\ds$ takes a state-dependent value $\ds^{\rm meso}$ comprised between 1 and 4 \cite{COT3}. We regard $\ds^{\rm meso}$ as the not-too-deep UV value of the spectral dimension in \Eq{gnude}, since LQC models of effective dynamics are not valid in the discreteness regime. Therefore, evaluating the left-most hand side of \Eq{lqcd} in this regime we would roughly get $\dh^{\rm meso}-\tilde{d}_{\rm W}^{\rm meso}-2\simeq-4/\ds^{\rm meso}$, which contradicts the right-hand side of \Eq{lqcd} because $\ds^{\rm meso}$ is non-negative. This is not completely surprising, since the cosmological model \Eq{mukkh} is semi-classical and is not expected to hold in the short mesoscopic scales of the states considered in \cite{COT3}. Therefore, we can look at an intermediate regime not too far from the IR, where numerical simulations show that the Hausdorff dimension tends to 4 from below, while the spectral dimension tends to 4 from above after crossing a local maximum \cite{COT3}. We can parametrize these near-IR profiles as $\dh^{\rm near-IR}=4-\e_{\rm H}$ and $\ds^{\rm near-IR}=4+\e_{\rm S}$, where $\e_{\rm H}$ and $\e_{\rm S}$ are positive. Moreover, typically the Hausdorff dimension runs to 4 faster than the spectral dimension \cite{COT3} and we can safely assume that $\e_{\rm S}>\e_{\rm H}>0$. Then, the left-most hand side of \Eq{lqcd} reads
\be
\dh^{\rm near-IR}-\tilde{d}_{\rm W}^{\rm near-IR}-2=\frac{\e_{\rm S}-\e_{\rm H}}{2}>0\,,
\ee
consistently with \Eq{lqcd}. Therefore, we interpret the right-most hand side \Eq{lqcd} as the difference in the running of the spectral and Hausdorff dimensions near the IR. In this regime, $\g=1+(\e_{\rm S}-\e_{\rm H})/4\simeq 1$, $h\sim 1/d_L^\textsc{gw}$ and
\be\label{de2lqc}
\Xi=\frac{a}{a_{\rm inv}}\simeq 1+\frac{\a_0}{2}\dpl(z)=1+\frac{\a_0}{2}\,\left(\frac{a_*}{a_0}\right)^\s(1+z)^{\s}\,,
\ee
where we keep the value $a_0$ of the scale factor today generic. This expression is valid for not too large redshifts, including $z=O(1)$, and is compatible with the parametrization \Eq{eq:param}.

The anomaly-cancellation LQC model with holonomy corrections predicts \cite{BH2,CMBG,CBGV}
\be\label{mued}
w_k''+\left(s^2_{\rm hol}k^2-\frac{a_{\rm hol}''}{a_{\rm hol}}\right)w_k=0\,,
\ee
where
\be\label{}
s^2_{\rm hol} =1-2\frac{\rho}{\rho_*}\,,\qquad a_{\rm hol}:=\frac{a}{|s_{\rm hol}|} \,,
\ee
and the energy density $\rho_*$ is about 0.42 the Planck energy density. Therefore,
\be\label{de2lqc2}
\Xi=|s_{\rm hol}|=\frac{a}{a_{\rm hol}}= \sqrt{1-2\frac{\rho(z)}{\rho_*}}\simeq 1-\frac{\rho(z)}{\rho_*}\geq 0\,.
\ee
Also this expression is compatible with the parametrization \Eq{eq:param}, after expanding in $1+z$ (the energy density is a function of $a$, hence of $1+z$). The tensor spectrum is blue tilted also in this model. The model with a quadratic inflaton potential is in tension with \textsc{Planck} data unless one fine tunes the initial conditions in the inflaton field \cite{BoBGS}.

In all these cases, one should consider also constraints from the propagation speed $s_{\rm GW}$ of gravitational waves (the coefficient in front of the $k^2$ term in the Mukhanov--Sasaki equation). The joint analysis by LIGO-Virgo and \emph{Fermi} of the gravitational-wave event GW170817 and its optical counterpart (gamma-ray burst) GRB170817A yielded the stringent upper bound $s_{\rm GW}-1<7\times 10^{-16}$ in $c=1$ units \cite{Ab17b}. From \Eq{mukkh}, the propagation speed in the effective-dynamics LQC model with inverse-volume corrections is $s_{\rm GW}\simeq 1+\a_0\dpl$, so that
\be
\a_0\dpl<10^{-15}\,.
\ee
Taken alone, this constraint states that the LQC inverse-volume quantum correction is negligible in the velocity of observed GWs emitted from these astrophysical sources, although its magnitude can be much larger at the epoch of production of the cosmic microwave background \cite{BCT2,Zhu15,Zhu16}. However, thanks to equation \Eq{de2lqc} we reach an even stronger conclusion and rule out the model also as a source of modification to the luminosity distance.

Similarly, in the LQC model with holonomy corrections $s_{\rm GW}=|s_{\rm hol}|\simeq 1-\rho/\rho_*<10^{-15}$, which comes as no surprise since $\rho/\rho_*\ll 1$ at late times. Thus, from \Eq{de2lqc2} $\Xi\approx 1$.

To summarize, in the effective-dynamics approach of LQC quantum corrections are not obviously small (at least, the inverse-volume ones) because they depend on the details of the underlying quantum state of geometry. We excluded that model as an interesting candidate only when combining luminosity-distance observations with constraints on the propagation speed $s_{\rm GW}$ of the tensor perturbation. It may be that other LQC scenarios, such as the hybrid-quantization approach \cite{lqcr,EMM}, could evade speed constraints (because $s_{\rm GW}=1$ there) and, at the same time, admit a quantum state with sizable IR corrections.
	

\subsection{Other models}\label{oth}

We will now briefly mention some other cosmological scenarios inspired by quantum gravity, that may have an impact in GW astronomy.
These models will, however, require more study in order to extract the observables.

\subsubsection{Non-commutative cosmology}\label{ncc}

In the Brandenberger--Ho class of models, inflation is realized by a scalar field evolving in a four-dimensional spacetime with a non-commutative structure 
\cite{BH}. This structure is determined by a spacetime uncertainty relation that could emerge, among other possibilities, from string theory. In a cosmological setting, calling $\eta =\int a\,dt=\int da/H$ (not to be confused with conformal time), the uncertainty relation is $\Delta \eta \Delta x \geq \ell_*^2$, where $\ell_*$ is a fundamental scale. This uncertainty relation can be written as a commutator between time and space coordinates
\be \label{alg}
[\eta,x]=\rmi \ell_*^2\,.
\ee

Two of these models, belonging to the so-called class 1 proposal in the strongly non-commutative IR regime ($\ell_* H\ll 1$, modes generated outside the horizon) \cite{BH,Cal4}, have a blue-tilted tensor spectrum \cite{Cal4} and are compatible with \textsc{Planck} data when a natural-inflation potential is assumed \cite{CKOT}. Unfortunately, these models cannot give rise to a detectable primordial stochastic signal at LISA frequencies because the IR regime affects only very large scales, while shorter scales enter a UV regime ($\ell_* H\gg 1$, modes generated inside the horizon) with a red-tilted tensor spectrum. Also models in the UV regime are viable in the early universe \cite{CKOT} and, although they do not generate a stochastic GW background interesting for LISA, we may wonder whether they can affect the propagation of gravitational waves emitted at a later stage by astrophysical systems.

In these scenarios, 
 the Mukhanov--Sasaki equation for tensor perturbations is \cite{BH,Cal4}
\be\label{nc}
\p_{\tilde\tau}^2w_k+\left(k^2-\frac{\p_{\tilde\tau}^2a_{\rm eff}}{a_{\rm eff}}\right)w_k=0\,.
\ee
Here $\rmd\tilde\tau=a^{-2}_{\rm eff}\rmd\eta$ and the effective scale factor $a_{\rm eff}$ is made of the scale factor $a(\eta)$ and the ``smeared'' functions $a_\pm:= a(\eta\pm k\ell_*^2)$. For instance, in the above-mentioned class 1 models, $a_{\rm eff}=\sqrt{a_+ a_-}$ \cite{BH} or $a_{\rm eff}=\sqrt{a}(a_+ a_-)^{1/4}$ \cite{Cal4}. Therefore, 
\be
\Xi(z,k)=\frac{a(z)}{a_{\rm eff}(z,k)}
\ee
is a function of $z$ and of $k$ and Nishizawa's parametrization \cite{Nis17} may be more suitable than \Eq{eq:param} in this case.

One may also consider the gravitational sector of the bosonic part of the non-commutative spectral action, an approach that provides a purely geometric interpretation of the Standard Model of strong and electroweak interactions \cite{Chamseddine:2006ep}.
By studying the propagation of gravitational waves \cite{Nelson:2010rt}, one may constrain \cite{Nelson:2010ru,Lambiase:2013dai,Capozziello:2014mea} the free parameter of the model, which is physically related to the coupling constants at unification. On the other hand, the generation of GWs can be affected by the scale of non-commutativity in non-commutative spacetimes, which is constrained to be around the Planck size \cite{Kobakhidze:2016cqh}.

\subsubsection{Varying-speed-of-light models}\label{vsl}

In varying-speed-of-light models \cite{Mag00,Mag03}, the speed of light is spacetime-dependent and local Lorentz invariance is preserved when using the time coordinate $x^0=\int\rmd t\, c(t)$. Although early models are ruled out, recent ones related to bimetric gravity and doubly special relativity can explain the observed cosmic structure \cite{Mag08a,Mag08b,Mag08c,Mag10}. In this case, the luminosity distance is \Eq{dL} with
\be
v(z)=c(z)
\ee
and the dynamics of $c(z)$ is encoded in the one of a scalar field $\chi:=\ln (c/c_0)$. One can trade a varying $c$ with a theory with a constant $c_0$ where gravity is non-minimally coupled to $\chi$. Here we do not extract the luminosity distance $d_L(z)$ predicted by varying-speed-of-light models but the interested reader should keep in mind this possibility.







\section{Strain noise and quantum gravity}\label{stn}

Quantum-gravity effects can be constrained by modern GW interferometers despite being the fundamental scale $\ell_*$ very small, even of order of the Planck scale \cite{Ame98,NgDa2,Ame13,Mar16,Bosso:2018ckz}. The constraint from \Eq{dla} can be complemented by a bound on $\dh^{\rm UV}$ coming from the strain noise of LISA. The correction in \Eq{L} can also be regarded as a threshold on the minimal uncertainty in physical measurements (spacetime fuzziness) of distances. We will review the argument of \cite{Ame98,Ame13}, recently revived in \cite{ACCR} in the context of multi-fractional theories, and apply it to LISA and other interferometers.\footnote{Depending on how they are implemented, other generic QG-inspired arguments based on fuzziness (Planck-scale quantum uncertainty) may lead to more pessimistic estimates of observable effects \cite{AMY}.}

In the stochastic interpretation of \Eq{L}, quantum gravity may manifest itself as an intrinsic noise with variance
\be\label{sqg}
\s_{\rm QG}^2=\ell_*^2\left(\frac{L}{\ell_*}\right)^{2\a}.
\ee
If $L$ is the typical length of a GW interferometer (for example, $L$ may be the linear size of its arms), we can compare this QG noise with the instrumental or strain noise $\s^2_{\rm exp}=\int\rmd f\,\cS^2(f)$ of a GW interferometer, where $f$ is the frequency of the signal (in Hz) and $\cS$ is the spectral noise. For a signal dominated by the characteristic frequency $c/L$, a rough estimate is $\s^2_{\rm exp}\simeq f\,\cS^2(f)|_{f=c/L}$. The strain noise is dimensionless and \Eq{sqg} has the dimensionality of (length)$^2$, so that a signal of spacetime fuzziness would be detectable if
\be\label{ss}
\left(\frac{\ell_*}{L}\right)^{2(1-\a)} = \frac{\s_{\rm QG}^2}{L^2} \sim \s^2_{\rm exp} \simeq f\,\cS^2(f)\big|_{f=\frac{c}{L}}\,.
\ee
This is the main argument of \cite{Ame98,Ame13}, where a number of purely phenomenological candidates (i.e., different values of $\a$) were proposed for the left-hand side. The main advance in \cite{ACCR} was to construct an explicit model motivated by quantum gravity generating the signal \Eq{sqg}, where $\a$ was recognized as a parameter related to the dimension of spacetime. Then \Eq{ss} was solved for $\cS$ and the solution inverted in $\a$ to find an upper bound on $\dh$ for LIGO. We redo this step to correct a typo in the final formula of \cite{ACCR}, which also affects the numerical result for LIGO. The left-hand side of \Eq{ss} is reproduced by the right-hand side with
\be\label{gS}
\cS(f)= \left(\frac{c}{\ell_*}\right)^{\a-1} f^{\frac12-\a}\qquad \Longrightarrow \qquad \a=\frac{\ln\left(\frac{c\,\cS}{\ell_*\sqrt{f}}\right)}{\ln\left(\frac{c}{\ell_* f}\right)}\,.
\ee
In the worse-case scenario, $\ell_*$ is of Planckian size, so small that one might believe it impossible to probe with an instrument of a macroscopic size $L$ of order of the kilometer (or millions of kilometers, in the case of LISA). However, $L$ does not appear in \Eq{gS} and, if $\a$ is small enough, the detector may even catch the stochastic background from spacetime fuzziness. Setting
$\ell_* =\lp\approx 1.6\times 10^{-35}\,{\rm m}$ in \Eq{gS}, we get an upper bound on $\a$ for all the main interferometers in operation, under construction, or proposed for the near future (Tab.\ \ref{tab4}).
\begin{table}
\centering
\begin{tabular}{l|c|c|c}\hline\hline
           & $\cS$ (Hz$^{-1/2}$) & $f$ (Hz) & $\a$   \\\hline
LIGO/Virgo/KAGRA & $10^{-23}$          & $10^2$   & $<0.47$ \\
LISA       & $10^{-20}$          & $10^{-2}$& $<0.54$ \\
DECIGO     & $10^{-23}$          & $10^{-1}$& $<0.47$ \\\hline\hline
\end{tabular}
\caption{\label{tab4} Upper bound on $\a$ from the strain noise.}
\end{table}

The constraint on $\a$ translates into a bound on the small-scale Hausdorff dimension of spacetime,
\be\label{dhuvbou}
\dh^{\rm UV}<1.9\,,
\ee
very close to the value for GFT/spin foams/LQG. This is what we get from present LIGO observations and, unfortunately, LISA is not expected to do better. In fact, the greater the strain sensitivity, the stronger the upper bound for a given frequency. 
 It is interesting to see how gravitational-wave astronomy can unravel the details of the deep spacetime structure even indirectly through instrumental noise information.


\section{Conclusions}\label{conc}

In this paper, we considered quantum-gravity scenarios modifying the effective gravitational dynamics while leaving the electromagnetic sector untouched. In complete theories that include all sectors such as string theory, as well in QG with non-local UV modifications of kinetic terms of all fields, this assumption breaks down but it carries no appreciable impact, since the QG effect is small and unobservable in GW observations (see \cite{BrCM} for the case of non-local gravity). Also group field theory may be regarded as a theory of everything, but its study is still at an embryonic stage and it is not known what the matter sector looks like. In other QG theories where gravity is quantized independently of a written-by-hand Standard Model, only the gravitational sector is modified and QG effects do not propagate onto photons or matter, if the coupling is minimal. While in yet other theories, we simply do not know whether and how the Standard Model is modified. Finally, in models such as those of section \ref{vsl} we do know that the electromagnetic sector is modified, but we have not worked out the luminosity distance yet.

Be as it may, letting photons propagate as usual we extracted physical observables and parameters of interest that can allow us to test theories of quantum gravity against present (LIGO, Virgo), upcoming (KAGRA) and future (LISA, DECIGO) gravitational-wave observations. Some of our results are based on concrete models, while others are of general applicability in any case admitting a cosmological continuum limit; some quantities were calculated only in the small-redshift limit (the range of LIGO, Virgo and KAGRA) while others are portable to any redshift and may be of relevance for LISA and DECIGO. Our analysis of deep-UV effects include QGs with a non-vanishing spectral dimension in the UV, while it does not apply to theories where $\ds=0$: the discreteness limit of group field theory, spin foams and loop quantum gravity (which we collectively dubbed as ``GFT/spin foams/LQG'') \cite{COT3}, the low-energy limit of string field theory and non-local quantum gravity \cite{Mod1,CaMo1}. These cases may deserve further attention. On the other hand, the analysis of near-IR effects applies to any theory in principle.

Most QG models do not give rise to observable modifications of GR, but the reason is not the same in all cases. We have seen that, in some cases, the quantum correction indeed takes the na\"ive form of a cosmological scale divided by the Planck scale. But in others the correction is not obviously small because it depends on the details of the underlying quantum state of geometry, which we described here in terms of geometric indicators such as the spectral and Hausdorff dimension. Even if more in-depth studies of specific models of non-perturbative theories confirmed these results, we do not expect many survivors that could give an IR signal. For instance, one could use the methods of \cite{COT3} to construct GFT/spin foam/LQG quantum states giving rise to a large $\Gamma_{\rm meso}$ and check whether these states are realistic or not. Allowing for the possibility that some QG models give modifications at large cosmological distances without contradicting any other cosmological observation may sound paradoxical, but one should consider the fact that known constraints on GFT/spin foam/LQG only come from inflationary scales and are of a rather different nature than those found here \cite{CQC}. Not only is the scaling dependence of the luminosity distance \Eq{dla} (length + power of length) typical of dimensional-flow effects, but, moreover, the power $\g$ in the correction is a combination of spacetime dimensions unique to propagating GWs. Also, spin foams and LQG do not modify the Standard Model and precision tests on the constants of Nature are preserved. \emph{A priori}, one might find a QG imprint at cosmological scales without violating particle and atomic physics, as the example of string theory reminds us. Although solar-system tests can constrain the same parameter space as GWs and be more stringent, as we saw in section \ref{compa}, GW multi-messenger astronomy can yield information not available in solar-system observations, as the examples in sections \ref{lqc} and \ref{stn} illustrate. These results rely on the set of assumptions listed throughout the paper: unmodified electromagnetic sector, identification $h_{00}\propto\Phi$ in the same regime where the dynamics \Eq{hboxh2} holds, and absence of local screening mechanisms. Breaking even one of them could lead to significant departures that would deserve, in our opinion, to be studied in the future.

Additional constraints on the spin-2 sector, which we have not mentioned so far, can arise from GW production models and, in particular, from observations of the Hulse--Taylor pulsar \cite{WeTa}. If the spacetime dimension deviates from four roughly below scales $l_{\rm pulsar}=10^6\,{\rm km}\approx 10^{-13}\,{\rm Mpc}$, then the GW emission from this source is expected to be distinguishable from GR. However, it is difficult to analyze the binary dynamics and GW emission in higher-dimensional spacetimes \cite{CDL} and it is consequently more complicated to set bounds from binary pulsar systems. These investigations may be worth further attention.

Our main message is that dimensional flow, a non-perturbative effect common in all quantum-gravity theories such that the dimension of spacetime changes with the probed scale, could make top-down scenarios beyond GR manifest in GW astronomical observations as well as in the solar system. The chance is slim but it should be contrasted with the worse odds when considering only perturbative effects. For instance, in the effective field-theory approach to quantum gravity \cite{Do94a,Do94b,Wei09} high-energy modes are integrated out, the Newton potential receives corrections of the form \Eq{effdon} \cite{BBDH2,KhKi} and the emission of gravitational waves is affected \cite{JPS,CELM}, but these corrections are heavily suppressed. Similarly, perturbative modifications in dispersion relations are suppressed at least as $\cK=k^2(1+\lp k)$ and they are virtually unconstrained in GW observations \cite{EMNan,ArCa2}, as we also appreciated in our analysis. Nevertheless, GW and solar-system tests \emph{are} capable of enforcing strong bounds not only on quantum gravities \cite{revmu,MYW,YYP}, but also on dark-energy phenomenological scenarios \cite{EzZu,KaTs}. Here we have seen that corrections from dimensional-flow are in general less suppressed than perturbative effects, as in the Newton potential \Eq{Newpot} or in the dispersion relations $\cK=k^2[1-(\ell_* k)^{1-\a}]$ with $0<\a<1$ discussed in \cite{revmu}. Our results show that getting a detectable signal from quantum gravity is no easy feat. However, at the present stage of GW astronomy it is important to understand how the theory can make contact with ground-based and space-borne interferometric experiments, and we pursued this goal consistently. Focus on the large-redshift luminosity distance as a promising observable for standard sirens will complement that on quantities based on local (solar system, supernov\ae, cosmic structures) observations to refine our understanding of which theories beyond GR will have the lion's share in the cosmology of the near future.


\section*{Acknowledgments}

\noindent The authors thank for discussions the members of the LISA Consortium Cosmology Working Group, where this project was first developed. G.C.\ thanks M.\ Arzano, A.\ Eichhorn, D.\ Oriti and J.\ Th\"urigen for useful email correspondence. G.C.\ and S.K.\ are supported by the I+D grant FIS2017-86497-C2-2-P of the Spanish Ministry of Science, Innovation and Universities. S.K.\ is supported by JSPS KAKENHI No.~17K14282 and Career Development Project for Researchers of Allied Universities. M.S.\ is supported in part by the Science and Technology Facility Council (STFC), United Kingdom, under the research grant ST/P000258/1. G.T.\ is partially supported by STFC grant ST/P00055X/1.

\appendix


\section{Anomalous scaling of the Green function}\label{app1}

In this Appendix, we find the GW amplitude associated with \Eq{hboxh2} for a particular model and in the local wave zone. This model assumes $v(x)=1$ in \Eq{hboxh2} and $w(k)=1$ in momentum space, i.e., trivial measure weights in position and momentum spaces. Therefore, $\dh=\dh^k=D$ here. Also, as a kinetic term we take the one giving rise to the modified dispersion relation \Eq{dr1}. Under these assumptions,
\be\label{gsta}
\G_*=\frac{D}{2}-\b\,.
\ee

Adapting the discussion in section \ref{gwalw}, the most general solution for $h_{ij}$ in the presence of a source and in the local wave zone (cosmic expansion neglected) is the generalization of \Eq{Gret} to the case of the kinetic operator $\cK=\ell_*^{2-2\b}\B+(-\B)^\b$. The solution is \Eq{scw},
\ba
G(x-x')&=&\frac{1}{(2\pi)^D}\int\rmd^Dk\,\rme^{\rmi k\cdot (x-x')}\frac{1}{\ell_*^{2-2\b}k^2-k^{2\b}}\nonumber\\
&=&\int\frac{\rmd\om}{2\pi}\,\rme^{-\rmi \om(t-t')}\int\frac{\rmd^{D-1}{\bf k}}{(2\pi)^{D-1}}\frac{\rme^{\rmi {\bf k}\cdot ({\bf x}-{\bf x'})}}{\ell_*^{2-2\b}({\bf k}^2-\om^2)-({\bf k}^2-\om^2)^\b}\,.\label{scw2}
\ea
Here we are interested in the scaling of the Green function, so we do not need to restrict our attention to the retarded one. The integral over spatial momenta is $\int\rmd^{D-1}{\bf k}=\int\Om_{D-1}\int_0^{+\infty}\rmd k\,k^{D-2}$, where $\Om_{D-1}$ is the angular integral and $k=|{\bf k}|$.
Choosing a frame where ${\bf k}\cdot ({\bf x}-{\bf x'})=kr\cos\theta$ and $r=|{\bf x}-{\bf x'}|$, one has
\ba
\int\frac{\rmd\Om_{D-1}}{(2\pi)^{D-1}}\,\rme^{\rmi {\bf k}\cdot ({\bf x}-{\bf x'})} &=& 
\frac{1}{2^{D-2}\pi^{\frac{D}{2}}\Gamma\left(\frac{D}{2}-1\right)}\int_0^\pi\rmd\theta (\sin\theta)^{D-3}\rme^{\rmi k r\cos\theta}\nonumber\\
&=& \frac{1}{(2\pi)^{\frac{D-1}{2}}}\frac{1}{(kr)^{\frac{D-3}{2}}}J_{\frac{D-3}{2}}(kr)\,,\label{bes}
\ea
where $J_\nu$ is the Bessel function $J$ of the first kind\index{Bessel functions} of order $\nu$ and we used \cite[formula 3.915.5]{GR}. 

To solve \Eq{scw2} analytically, we take a monomial dispersion relation, corresponding to a regime with $\G\approx {\rm const}$. Setting $t'=0$ and using \Eq{bes} and the analytic continuation of \cite[formula 6.565.4]{GR},
\ba
G(t,r)&\simeq&-\frac{1}{(2\pi)^{\frac{D+1}{2}}}\frac{1}{r^{\frac{D-3}{2}}}\int_{-\infty}^{+\infty}\rmd\om\,\rme^{-\rmi \om t}\int_0^{+\infty}\rmd k\,\frac{k^{\frac{D-1}{2}}J_{\frac{D-3}{2}}(kr)}{({\bf k}^2-\om^2)^\b}\nonumber\\
&=&-\frac{1}{(2\pi)^{\frac{D+1}{2}}}\frac{1}{r^{\frac{D-1}{2}-\b}}\int_{-\infty}^{+\infty}{\rmd\om}\,\rme^{-\rmi \om t}\frac{(\rmi\om)^{\frac{D-1}{2}-\b}}{2^{\b-1}\G(\b)}\,K_{\frac{D-1}{2}-\b}(\rmi\om r)\,.
\ea
In the local wave zone approximation \Eq{lwfa} $\om r\gg 1$, $K_\nu(\rmi\om r)\sim (\om r)^{-1/2}$ up to a phase, so that
\ba
G(t,r)&\sim& \frac{1}{r^{\frac{D}{2}-\b}}\int_{-\infty}^{+\infty}{\rmd\om}\,\rme^{-\rmi \om t}\om^{\frac{D}{2}-1-\b}\nonumber\\
&=&\frac{f(t)}{r^{\frac{D}{2}-\b}}\propto \frac{1}{r^{\G_*}}\,,
\ea
where we used \Eq{gsta}. The result agrees with \Eq{Newton} after convoluting with the source as in \Eq{sou}.


\end{document}